%% file: discalchep.tex
\nopagenumbers

\input onetime

\input epsf

\input tables.tex
\noblackbox
\nref\one{M. Aganagic, A. Klemm, and C. Vafa, hep-th/0105045.}
\nref\two{D. Morrison and M. Plesser, Nucl. Phys. {\bf B440} (1995), 
hep-th/9412236.}

\def\tx{\vbox{\sl\centerline{Physics Department}%
\centerline{University of Texas at Austin}%
\centerline{Austin, TX 78712 USA}}}

\Title{\vbox{\baselineskip12pt
\hbox{UTTG-18-01}\hbox{hep-th/0112039}}}
{\vbox{\centerline{Calculation of Nonperturbative Terms}\centerline
{in Open String Models}}}

{\bigskip
\centerline{Julie D. Blum}
\bigskip
\tx

\bigskip
\medskip
\centerline{\bf Abstract}
Nonperturbative corrections in type II string theory corresponding to Riemann
surfaces with one boundary are calculated in several noncompact
geometries of desingularized orbifolds.  One of these models has a 
complicated phase structure which is explored.  A general condition
for integrality of the numerical invariants is discussed.

\Date{12/01}

\newsec{Introduction}

String theory has provided mathematicians with many interesting conjectural
results
that would be in some cases significantly more difficult to obtain 
through traditional techniques.  Calculations that count the number of maps
from Riemann surfaces into Calabi-Yau manifolds are one example.  Recently,
these calculations have been extended to Riemann surfaces with boundaries.
The addition of boundaries for a generic type II string theory reduces 
the supersymmetry to $N=1$ in four dimensions, and the counting of maps
corresponds to holomorphic terms in the field theory.   Such 
calculations are possibly relevant to an extension of the Standard Model.

In the following we will calculate the nonperturbative terms in a type II 
string theory generated by Riemann surfaces with one boundary, ``disk
instantons''.  Most of the techniques that will be employed here are
discussed in \one , \two , and other places.  
In honor of human decency, 
no additional references will be
mentioned.  We will focus on 
three noncompact models, the blowups of the
${\bf Z}_2 \times {\bf Z}_2$, ${\bf Z}_2 \times {\bf Z}_4$, and ${\bf Z}_7$
orbifolds of ${\bf C}^3$.  The three models support the 
conjecture that numerical invariants related mathematically to the Euler
characteristic of the moduli space of open string instantons and 
physically to the counting of domain walls are integral in phases where
Kahler parameters have a geometric interpretation.  The second
model has an intricate moduli space with many phases.  We explore the
phase structure of this model and see how it reduces in various limits to
simpler models.  The complications of this model necessitated an
understanding of what conditions ensure integrality of the numerical
invariants.  We explain why the geometric phases always give integers and
the conditions under which fractions are possible.

\newsec{${\bf Z}_2 \times {\bf Z}_2$}

\subsec{Toric Geometry}

The toric geometry for this model can be described by a linear sigma model
(two dimensional abelian gauge theory with $N=2$ supersymmetry).  There are 
six chiral fields carrying the following charges under three $U(1)$ gauge
fields.

\eqn\one{\eqalign{l_1&=(1,0,0,1,-1,-1)\cr l_2&=(0,1,0,-1,1,-1)\cr 
l_3&=(0,0,1,-1,-1,1)\cr}}
The Calabi-Yau condition requires $\sum_j{l_i^j}=0$.
From the set of charges, one derives the D-terms in the gauge theory.
\eqn\two{\sum_j {l_i^j |x_j|^2} =r_i}
These equations can be solved leading to the following ``toric diagram''.

\centerline{\epsfxsize=0.5\hsize\epsfbox{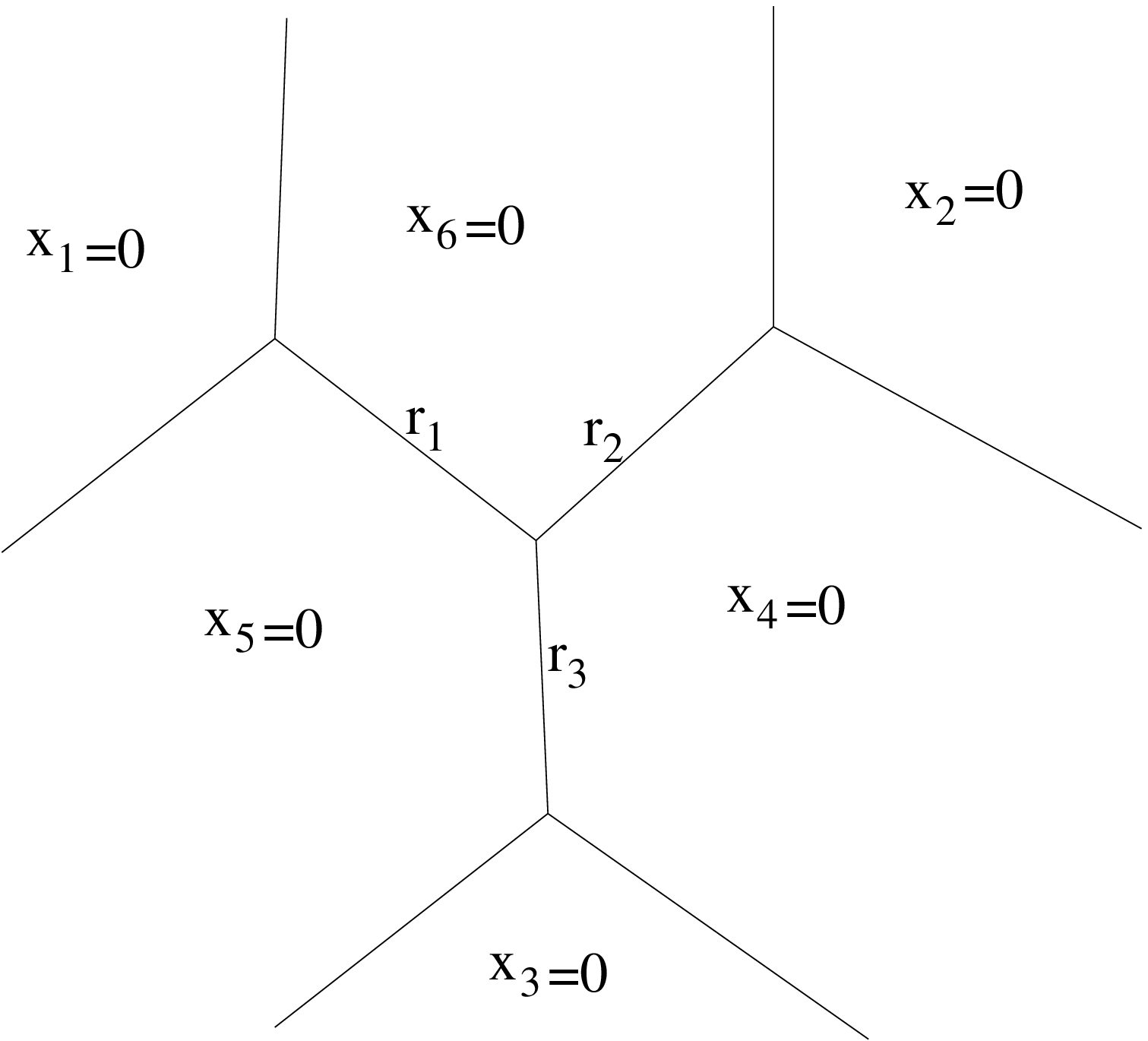}}
\bigskip
\centerline{\vbox{\noindent{\bf Diag. 1.}
Toric Diagram of ${\bf Z}_2\times{\bf Z}_2$ Blowup
}}
\vskip .7cm

The diagram is a projection of three dimensions onto the plane, and the 
angles shown are not meant to be accurate.  Generically, the diagram 
represents a ${\bf T}^3$ fibration.  It shrinks to a two-torus along planes
$x_i=0$, to a circle along lines $x_i=x_j=0$, and to a point at the
intersection of three lines.  From the diagram one sees that there are 
three two-spheres and six noncompact two-cycles.  Since this
diagram resembles the intersection of three conifolds, one would expect
other phases that replace a ${\bf P}^1$ by an ${\bf S}^3$.  One 
can easily visualize a
flopped phase as the following where $r_3\rightarrow -r_3$. 

\centerline{\epsfxsize=0.5\hsize\epsfbox{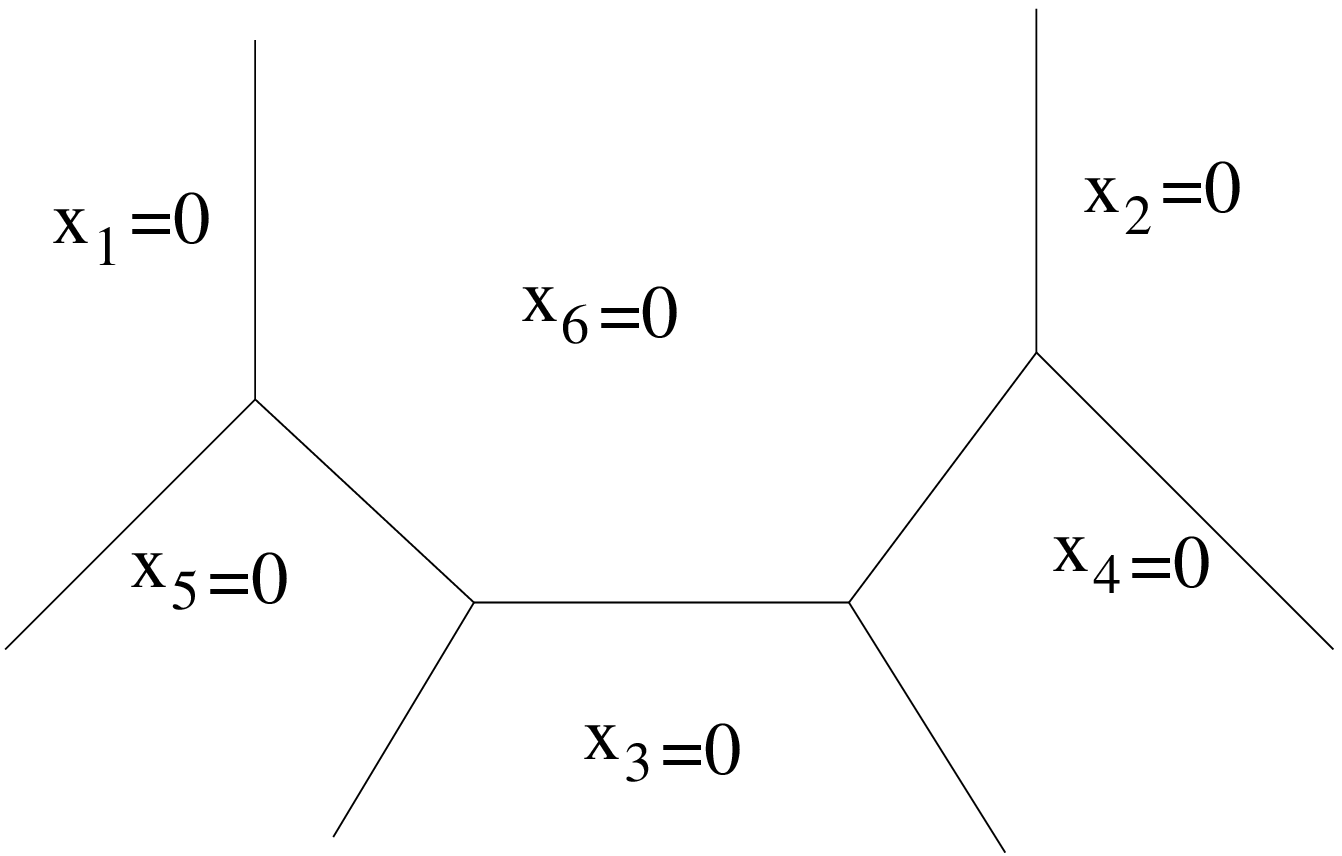}}
\bigskip
\centerline{\vbox{\noindent{\bf Diag. 2.}
Toric Diagram of ${\bf Z}_2\times{\bf Z}_2$ Flopped Phase:$r_3\rightarrow -r_3$
}}
\vskip .7cm

The equation 
for the
${\bf Z}_2 \times {\bf Z}_2$ singularity can be obtained from gauge
invariant combinations of chiral fields as $w^2=xyz$ where $w$, $x$,
$y$, and $z$ are complex variables.

To show explicitly that there are three ${\bf P}^1$'s, one should solve the 
Picard-Fuchs equations.  These equations determine the Kahler parameters
for the  ${\bf P}^1$'s corrected by worldsheet instantons or equivalently 
complex parameters in the local mirror geometry, i.e. 
$\int_{\gamma_i}
\Omega$, where $\Omega$ is the holomorphic three-form and $\gamma_i$ is a
three-cycle.  For this model the equations are
\eqn\three{[\Theta_1(\Theta_1 -\Theta_2-\Theta_3)-z_1(\Theta_1^2 -(\Theta_2-
\Theta_3)^2)]\int_{\gamma_i}
\Omega=0}
and cyclic permutations where $\Theta_i=z_i\partial_{z_i}$ and 
$z_i=e^{-t_i}$ with
$t_i=r_i +i\theta_i$, the initial Kahler parameter.  Here $\theta_i$ is a
Fayet-Iliopoulos parameter in the linear sigma model.  One takes linear
combinations of the $\int_{\gamma_i}
\Omega$ normalized appropriately to obtain ${\hat t}_i$, the instanton
corrected Kahler parameters.  For the calculation 
here we need the inverse solutions which turn out to be
\eqn\four{z_1={q_1(1+q_2 q_3)\over (1+q_1 q_2)(1+q_1 q_3)}}
and permutations where $q_i=e^{-{\hat t}_i}$.  Note that in finding unique 
solutions for the Picard-Fuchs
equations, we have frequently had to change to a different basis of 
${\bf P}^1$'s.
The simplicity of these solutions makes this model amenable to obtaining
exact results without great labor.

\subsec{The Mirror and Open String Amplitudes}

The equation for the mirror is readily derived from $Re(y_i)=-|x_i|^2$, the
D-term equations, and $xz=\sum_i{e^{y_i}}$ where $x$ and $z$ are
complex variables.  Setting $y_5=u$, $y_6=v$, and
$y_4=0$ to fix a constant solution of the D-term equations yields
\eqn\five{xz=P(u,v)=1+e^u+e^v+e^{-t_1 +u+v}+e^{-t_2 +v-u}+e^{-t_3 +u-v}.}
A noncompact, supersymmetric Lagrangian three-cycle in the original 
manifold is determined
by three additional constraints which in this case take the form 
\eqn\six{\eqalign{|x_5|^2 -|x_4|^2&=c_1\cr |x_6|^2 -|x_4|^2&=c_2\cr \sum_i
{Arg(x_i)}&=0\, \rm{and/or}\, \pi .\cr}}
Let us write the above in the form 
\eqn\laterone{\sum_j l_p^{(1) j}|x_j|^2|_{3-cycle}=c_p}
and 
\eqn\latertwo{\sum_j l^{(2)i}Im(log (x_i|_{3-cycle}))=0\, {\rm and/or}\, \pi}
for later use.  Here $\sum_i l_p^{(1) i} l^{(2)i}=0$
makes the cycle Lagrangian and $\sum_i l_p^{(1) i}=0$ is necessary for 
supersymmetry of the Lagrangian cycle.  Additionally, $\sum_i l_j^i l^{(2)i}=0$
implies that \latertwo\ is gauge invariant.
If the cycle does not intersect a compact or noncompact two-cycle in the
base, both $\sum_i{Arg (x_i)}=0$ and $\sum_i{Arg (x_i)}=\pi$ are needed to give
a composite cycle without boundary.  For this case any worldsheet disks
that intersect a D-brane wrapped on the two cycles will be oppositely
oriented with respect to the two cycles and not make a contribution.  If
the cycle intersects the toric base, one can choose either 
$\sum_i{Arg (x_i)}=0$ or $\sum_i{Arg (x_i)}=\pi$ to get a cycle without 
boundary, and there generally will be a nonvanishing contribution from 
disks wrapping part of a ${\bf P}^1$ and intersecting the cycle in a circle.
Allowing the three-cycle to end on the ${\bf P}^1$ where $x_6=x_4=0$ (which 
will be denoted as Phase I), the 
classical limit in the mirror corresponds to 
$v=i\pi$, ${\rm Re} u=-c_1\approx {-r_2\over 2}\rightarrow -\infty$, and 
$xz=0$.  More generally, one can choose the mirror two-cycle of the 
Lagrangian three-cycle to be parametrized by $z$ with $x=P(u,v)=0$, a
Riemann surface which is the moduli space of this cycle.  The coordinates
$u$ and $v$ can be considered as transverse coordinates to a two-cycle
inside a Calabi-Yau manifold.  

The disk amplitude ${\cal F}_{g=0,h=1}$ 
($g$ is the genus, $h$ is the number of boundaries) can
be determined classically in the mirror as $\partial_u {\cal F}_{0,1}=v$ 
where classically 
$v=0$ and $u$ parametrizes the area of a disk.  In the
original manifold, these disks can be interpreted as domain walls (e.g.
fourbranes wrapping a disk) ending on a sixbrane wrapped on the 
three-cycle.  The tension of these domain walls is corrected by an amount
$\delta u$ that must be added to the classical area of a disk.  The 
amplitude ${\cal F}_{0,1}$ takes the form
\eqn\seven{{\cal F}_{0,1}=\sum_{k,n,\vec m}{{1\over n^2 }N_{k,\vec m}
(\prod_i{q_i^{m_i n}})e^{{\hat u}kn}}}
where $k$, $n$, $m_i$ are integers, ${\hat u}$ is the instanton corrected
domain wall tension, and $N_{k,\vec m}$ counts the number of domain
walls wrapping the two-cycle parametrized by $\sum_i{m_i {\hat t}_i}$ 
with a boundary wrapping
the ${\bf S}^1$ $k$ times.  The assumption is that one counts isolated 
domain walls and that the $N_{k,\vec m}$ should be 
integers.  If one has a continuous family of domain walls, 
fractions may be possible.  In section three we will determine a 
criterion for obtaining integers and present an argument for 
integrality in those cases.  Under mirror symmetry a domain wall 
fourbrane becomes a domain
wall fivebrane with tension determined classically by ${\hat u}=u-\delta u$
where $\delta u ={1\over 2\pi i}\int_{C_u}u dv$ and $v\rightarrow v+2\pi i$
around the one-cycle on the Riemann surface $C_u$.  This tension 
corresponds to the difference in the superpotential ${\cal F}_{0,1}$
as one crosses the domain wall.  In the original manifold, the change in 
${\rm Im} v$ corresponded to the change in Wilson line as one crossed the 
domain wall.

There is an ambiguity in ${\cal F}_{0,1}$ due to the possibility of
redefining the disk coordinate ${\hat u}\rightarrow{\hat u}+n{\hat v}$ 
where $n$ is an integer since $v=0$ classically.  One requires $n$ to be
an integer so that $e^{\hat u}$ is invariant under ${\hat u}\rightarrow
{\hat u}+2\pi i$.  This ambiguity can sometimes be related to moving the 
Lagrangian cycle to a different phase along the toric base, and the amplitude
${\cal F}_{0,1}$ is not invariant.  If the toric base is modeled by
type IIB fivebranes, the ambiguity corresponds to an $SL(2,{\bf Z})$
transformation of type IIB.

Proceeding with the calculation, the Riemann surface $P(u,v)=0$ looks like
the following diagram where the legs extend to infinity. 

\centerline{\epsfxsize=0.5\hsize\epsfbox{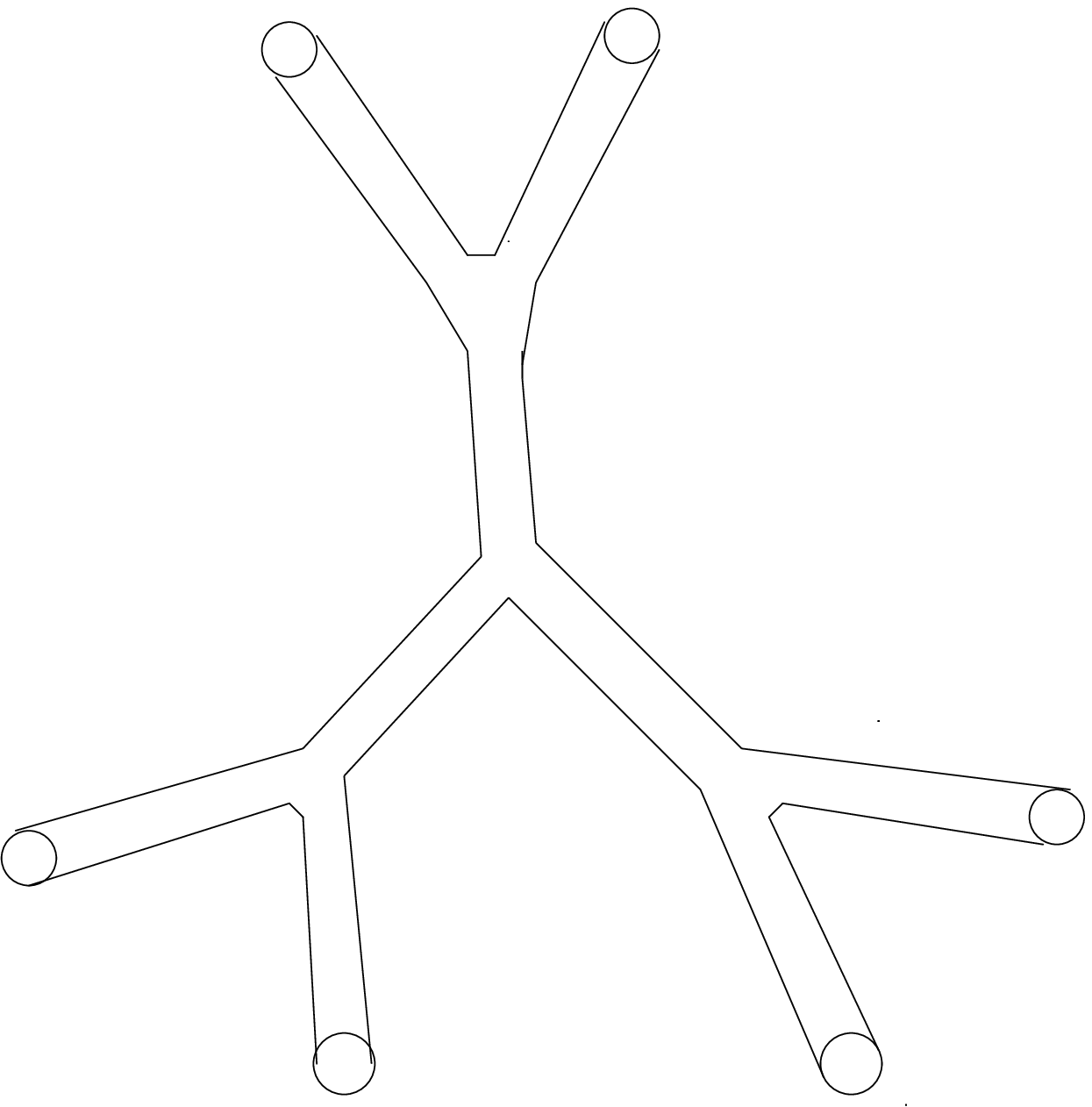}}
\bigskip
\centerline{\vbox{\noindent{\bf Diag. 3.}
Riemann Surface $P(u,v)=0$ in mirror of ${\bf Z}_2\times{\bf Z}_2$ Blowup
}}
\vskip .7cm

There are
nine phases, but the three compact ${\bf P}^1$'s and the
six noncompact two-cycles of the original manifold are related by
symmetries reducing the number of inequivalent phases to two.

\centerline{\epsfxsize=0.5\hsize\epsfbox{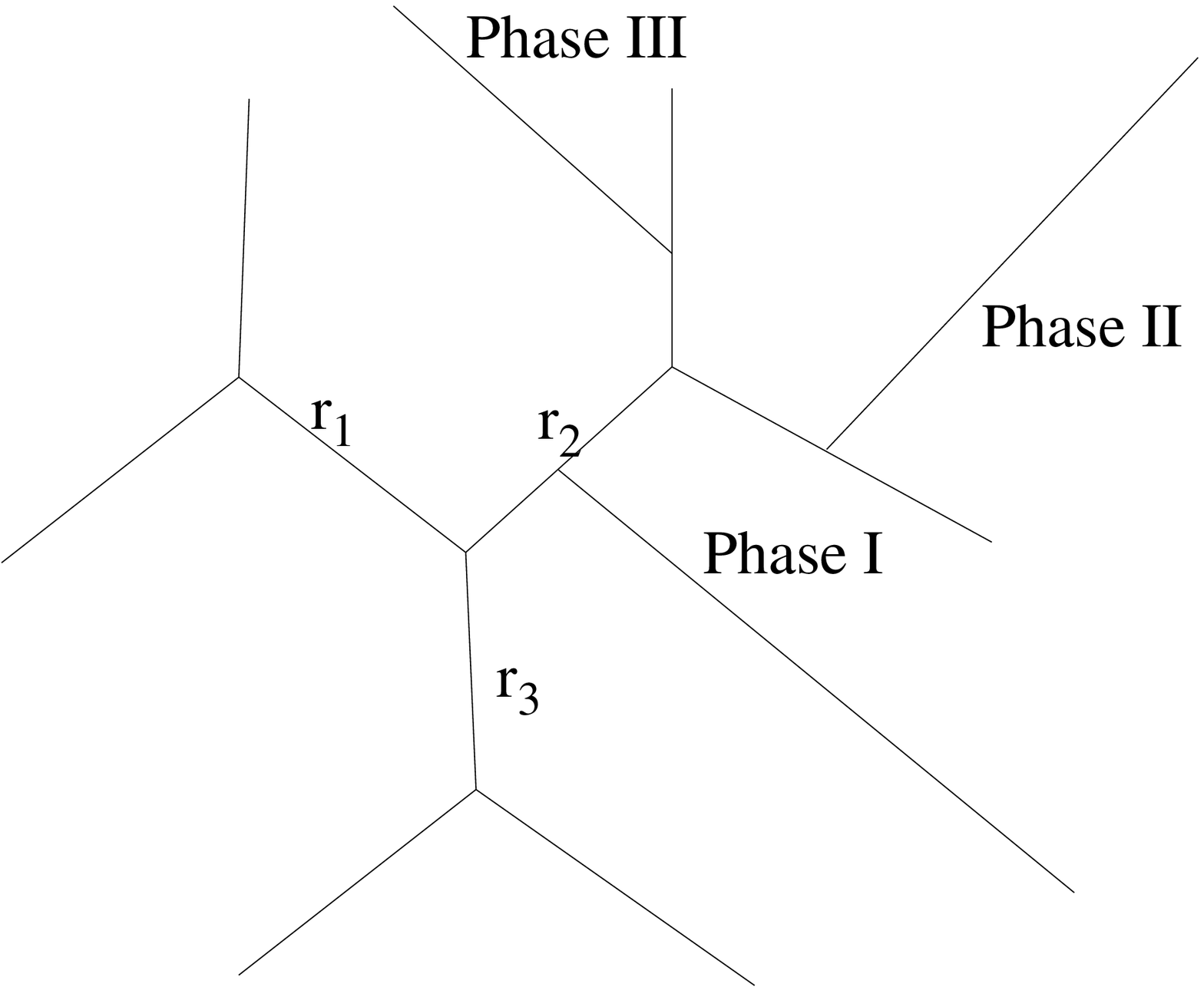}}
\bigskip
\centerline{\vbox{\noindent{\bf Diag. 4.}
Phases for noncompact three-cycle in ${\bf Z}_2\times{\bf Z}_2$ Blowup
}}
\vskip .7cm

For 
instance, the three inner
phases are exchanged by exchanging $z_3\leftrightarrow z_2\leftrightarrow
z_1$ along with $v\leftrightarrow u\leftrightarrow -u$, and phase II and 
phase III are exchanged under $z_1\leftrightarrow z_3$, $v\leftrightarrow -v$.
Phase II corresponds to $c_1=c_2+r_2$, $c_2>>0$.

In phase I one obtains
\eqn\eight{v=i\pi-\ln{2(1+z_1 e^u+z_2 e^{-u})}+\ln [(1+e^u)+\sqrt{(1+e^u)^2 -
4z_3(e^u +z_1 e^{2u}+z_2)}]}
Extracting the piece of $v$ that is independent of $e^u$ gives $\delta v$,
and one has 
\eqn\nine{\delta v ={t_1-{\hat t}_1\over 2}-{t_3-{\hat t}_3\over 2}+i\pi .}
To get $\delta u$, we exchange $t_3$ and $t_2$ so 
\eqn\ten{\delta u ={t_1-{\hat t}_1\over 2}-{t_2-{\hat t}_2\over 2}+i\pi .}
In the above equation for $v$, we must substitute, $v={\hat v}+\delta v$,
$u={\hat u}+\delta u$, and $z_i (q_i)$.  Note that not all of the classical
symmetries of the superpotential are preserved by the corrections, and one
cannot determine the correction uniquely by symmetry.
The result is 
\eqn\eleven{\eqalign{{\hat v}&=-\ln(1-q_1 e^{\hat u})(1-q_2 e^{-\hat u})\cr &-
\sum\nolimits '_{m,n,a,b,c,d,e}
{(-1)^{m+b+c}{(2n+m-1)!e^{{\hat u}(2a+b+c)}q_1^{a+d+e} q_2^{m+n-a-b-c+d} 
q_3^{m+n-c+e}
\over n! a! (c-e)! e! (b-d)!
(m-c)! d! (n-a-b)!}} \cr}}
The $\sum'$ indicates that we omit terms independent of $e^{\hat u}$.
Clearly the first term of ${\hat v}$ has the required form.  The lowest 
order terms of the second summation can be examined by hand or calculated on 
a computer using Mathematica, and one does obtain $N_{k,\vec m}$ that are
integers after integrating and comparing with \seven .  One can show 
explicitly that all $N_{1,\vec m}$ are integers.  One finds $N_{1,0,m,m}=-
\sum_{n=1}^m {{(-1)^{n+m}(n+m-1)! m\over n!^2(m-n)!}}$,
$N_{1,1,m-1,m}=\sum_{n=0}^{m-1} {{(-1)^{n+m}(n+m-1)! \over n!^2(m-n-1)!}}$,
and $N_{1,0,m-1,m}=N_{1,1,m,m}=\sum_{n=1}^m{(-1)^{n+m} (n+m-1)!\over n! (n-1)!
(m-n)!}$.  We have verified that all $N_{k,m_1 ,m_2 ,m_3}$ are integers for
$k\le 10$, $m_1\le 1$, $m_2\le 1$, and $m_3\le 10$.  Rather than present
this data which is very cumbersome, we give three tables 
with $k$ and $m_3$ set to
specific values.

\centerline{\bf{Table 1}:}
\smallskip
\centerline{$m_3=5$,\ $k=5$}
\smallskip
\begintable

$m_2$| $m_1$=$0$|$1$|$2$|$3$|$4$|$5$\cr
$0$|$5$|$-14$|$14$|$-6$|$1$|$0$\cr
$1$|$-126$|$350$|$-350$|$150$|$-25$|$1$\cr
$2$|$756$|$-2100$|$2100$|$-900$|$150$|$-6$\cr
$3$|$-1764$|$4900$|$-4900$|$2100$|$-350$|$14$\cr
$4$|$1764$|$-4900$|$4900$|$-2100$|$350$|$-14$\cr
$5$|$-635$|$1764$|$-1764$|$756$|$-126$|$5$
\endtable

\vskip 40pt
\centerline{\bf{Table 2}:}
\smallskip
\centerline{$m_3=5$,\ $k=6$}
\smallskip
\begintable

$m_2$|$m_1$=$0$|$1$|$2$|$3$|$4$|$5$\cr
$0$|$42$|$-126$|$140$|$-70$|$15$|$-1$\cr
$1$|$-630$|$1890$|$-2100$|$1050$|$-225$|$15$\cr
$2$|$2940$|$8820$|$9800$|$-4900$|$1050$|$-70$\cr
$3$|$-5880$|$17640$|$-19600$|$9800$|$-2100$|$140$\cr
$4$|$5292$|$-15876$|$17640$|$-8820$|$1890$|$-126$\cr
$5$|$-1764$|$5292$|$-5880$|$2940$|$-630$|$42$
\endtable
\vskip 20pt
The diagonal symmetry $N_{k,m_1-x,m_2-y,m_1+m_2}=N_{k,m_1+y,m_2+x,m_1+m_2}$
is generated by $d\leftrightarrow b-d$ in \eleven .  The similar symmetry 
$N_{k,m_1-x,m_1+m_3-k,m_3-y}=N_{k,m_1+y,m_1+m_3-k,m_3+x}$ is generated by
$e\leftrightarrow c-e$.

\centerline{\bf{Table 3}:}
\smallskip
\centerline{$m_3=5$,\ $k=7$}
\smallskip
\begintable

$m_2$|$m_1$=$0$|$1$|$2$|$3$|$4$|$5$\cr
$0$|$198$|$-630$|$756$|$-420$|$105$|$-9$\cr
$1$|$-2310$|$7350$|$-8820$|$4900$|$-1225$|$105$\cr
$2$|$9240$|$-29400$|$35280$|$-19600$|$4900$|$-420$\cr
$3$|$-16632$|$52920$|$-63504$|$35280$|$-8820$|$756$\cr
$4$|$13860$|$-44100$|$52920$|$-29400$|$7350$|$-630$\cr
$5$|$-4356$|$13860$|$-16632$|$9240$|$-2310$|$198$
\endtable

In phase II we do the coordinate transformation $u\rightarrow u'=u-v+t_2$, 
$v\rightarrow v'=v$.  Correspondingly, $\delta u'=\delta u-\delta v+\delta t_2
={t_3-{\hat t}_3\over 2}+{t_2-{\hat t}_2\over 2}$ and $\delta v'=\delta v$.
Phase II is almost equivalent to phase I.  One obtains phase II from
\eleven\ by ignoring the first log term, exchanging $q_2\leftrightarrow q_3$
and ${\hat u}\leftrightarrow {\hat v}$.  The first term is similar to
the inner phase of a conifold so it looks like the ${\bf S}_3$
symmetry relating the inner and two outer phases is broken by the
finite ${\bf P}^1$'s.

One can extract amplitudes in the flopped phase of Diagram(2) by taking
$q_3\rightarrow q_3 '=1/q'_3$, $q_1\rightarrow q'_1 q'_3$, $q_2
\rightarrow q'_2 q'_3$, and $\exp{\hat u}\rightarrow q'_3 \exp{\hat u'}$ in
\eleven .  
As a check on our results, the conifold
in the two inequivalent phases is retrieved in the limit $q_1,q_3\rightarrow
0$.  Replacing the ${\bf P}^1$ by an ${\bf S}^3$ in this limit via a
conifold transition also yields the same result from the calculation of
the expectation value of a Wilson line in the Chern-Simons theory on 
${\bf S}^3$.  It would be interesting to extend such calculations to
the case of multiple ${\bf S}^3$'s.

\newsec{${\bf Z}_2\times {\bf Z}_4$}

\subsec{Toric Geometry}

Let us move on to the ${\bf Z}_2\times {\bf Z}_4$ case.  Here we increase
the complexity of the calculation, but the results reduce precisely to
the ${\bf Z}_2\times {\bf Z}_2$ case in a particular limit.  We start
with the following set of charges under six $U(1)$'s for nine fields.
\eqn\thione{\eqalign{l_1&=(1,-1,0,0,0,-1,1,0,0)\cr l_2&=(0,-1,1,0,0,1,-1,0,0)
\cr l_3&=(0,1,0,0,0,-1,-1,0,1)\cr l_4&=(0,0,0,0,0,1,-2,1,0)\cr 
l_5&=(0,0,0,1,0,0,-1,-1,1)\cr l_6&=(0,0,0,-1,1,0,1,-1,0)\cr}}
yielding $\sum_j{l_i^j|x_j|^2}=r_i$.  Solving these equations leads to the
following toric diagram in one particular region of moduli space

\centerline{\epsfxsize=0.5\hsize\epsfbox{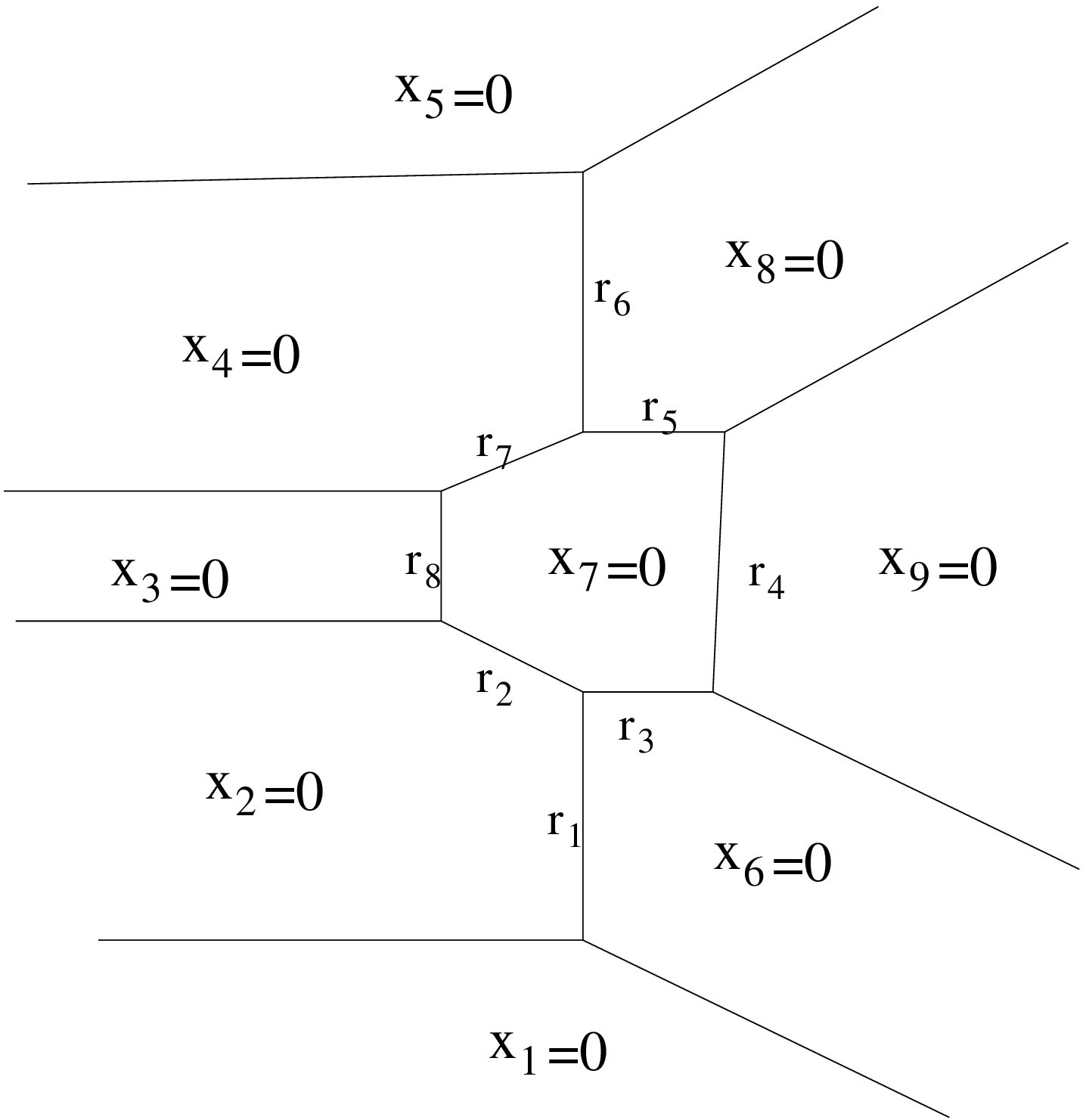}}
\bigskip
\centerline{\vbox{\noindent{\bf Diag. 5.}
Toric Diagram of ${\bf Z}_2\times{\bf Z}_4$ Blowup
}}
\vskip .7cm

\noindent where $r_7=r_2+r_3-r_5>0$, $r_8=-2r_2-r_3+r_4+r_5>0$, and
all of the $r_i$ are large.  
The four-cycle
represented by the hexagon is a ${\bf P}^2$ blown up in succession 
at three points (one obtains inequivalent four-cycles depending on 
how one does this).  It is also equivalent to the ${\bf F}_2$
Hirzebruch surface blown up at two points. 
By taking $r_4$ and $r_8$ to infinity, 
the above diagram and the theory reduces to two decoupled 
${\bf Z}_2\times {\bf Z}_2$ cases.  There are
possible flop transitions to other geometric phases for $r_1$, $r_2$, $r_3$, 
$r_5$, $r_6$, and $r_7$ but not for $r_4$ and $r_8$.  Shrinking $r_8$ to a 
negative value removes a ${\bf P}^1$ as $|x_3|>0$
everywhere, and we enter a nongeometric phase where a 
Kahler parameter loses its correspondence to a geometric ${\bf P}^1$.
The result of flopping $r_7$ is shown in Diagram (6).

\centerline{\epsfxsize=0.4\hsize\epsfbox{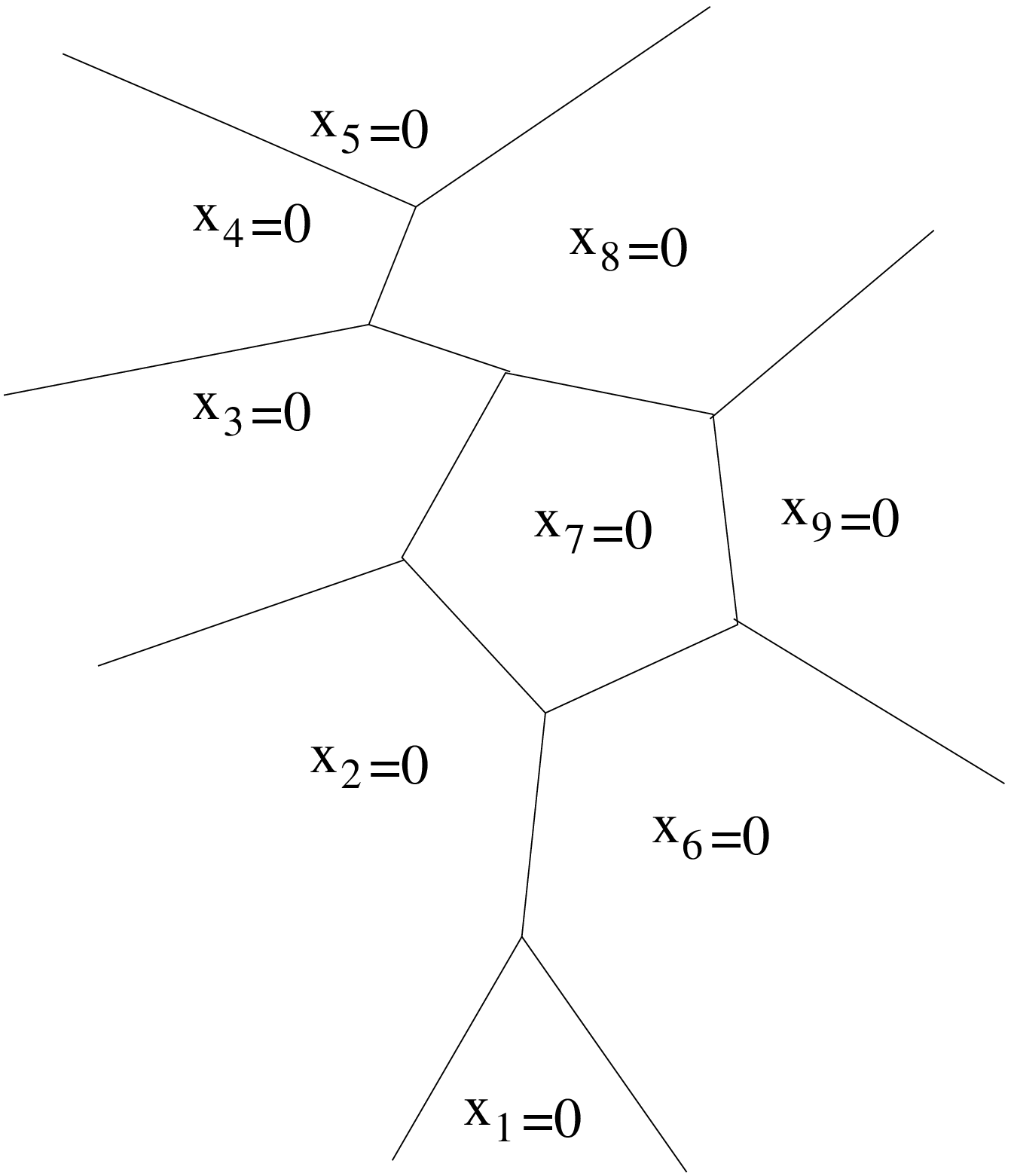}}
\bigskip
\centerline{\vbox{\noindent{\bf Diag. 6.}
Toric Diagram of ${\bf Z}_2\times{\bf Z}_4$ Flopped Phase:$r_7\rightarrow
-r_7$
}}
\vskip .7cm

\noindent Two more flops ($r_4 -r_2
\rightarrow r_2 -r_4$, $r_3\rightarrow -r_3$) and taking all $r_i$ to
infinity with $r_4-r_3>0$ and finite reduces the diagram to the blowup
of a ${\bf Z}_3$ orbifold.  The flops $r_2\rightarrow -r_2$ and 
$r_7\rightarrow
-r_7$ generate a ${\bf P}^1\times {\bf P}^1$, and one can take external
Kahler parameters to infinity to obtain this model.  The equations for the 
${\bf Z}_2\times {\bf Z}_4$ singularity are $v^2=yu$ and $v^4=wzu^2$.

\centerline{\epsfxsize=0.4\hsize\epsfbox{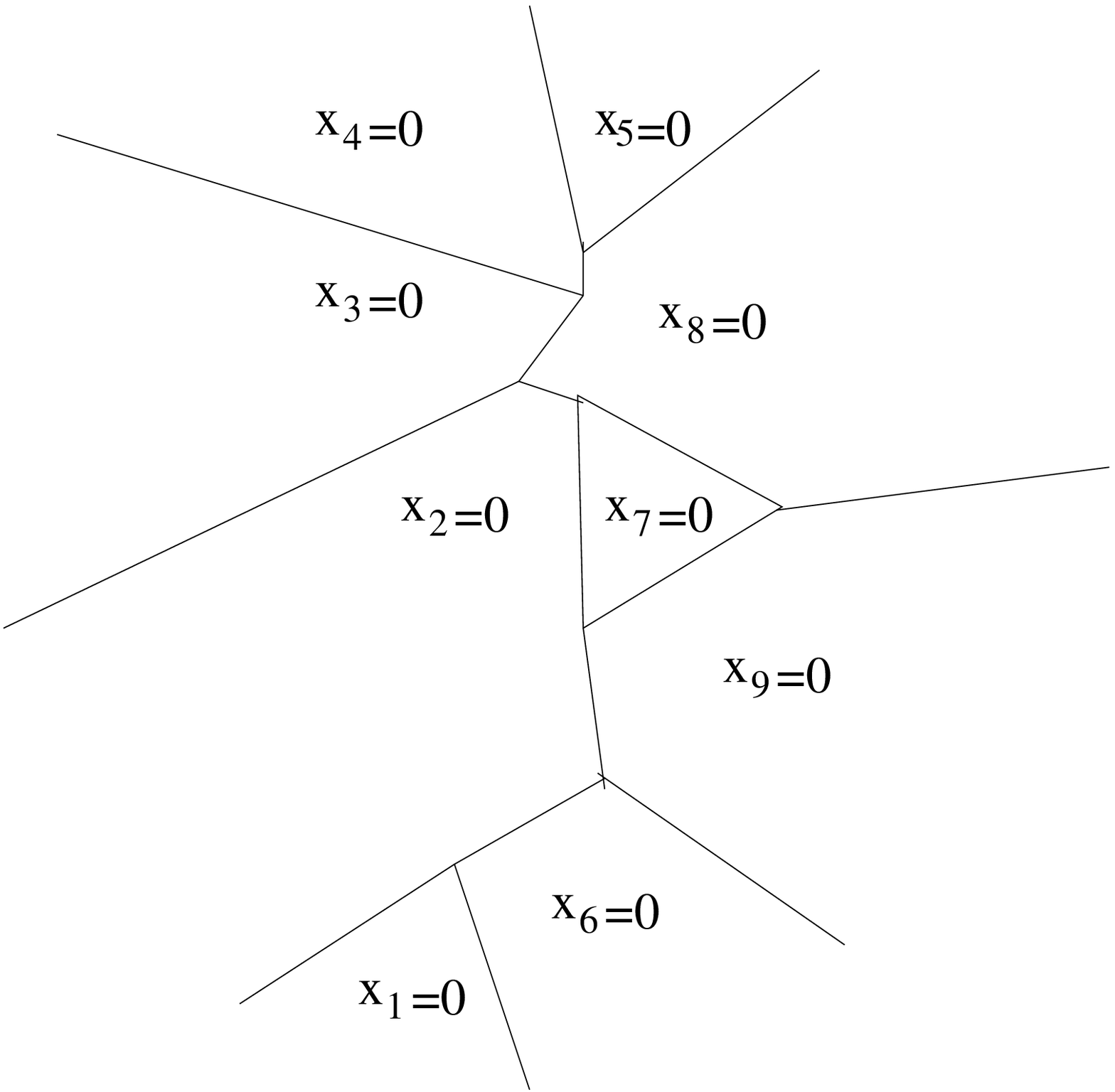}}
\bigskip
\centerline{\vbox{\noindent{\bf Diag. 7.}
Toric Diagram of ${\bf Z}_2\times{\bf Z}_4$ Flopped Phase:$r_7\rightarrow
-r_7$,$r_4 -r_2\rightarrow r_2 -r_4$,$r_3\rightarrow -r_3$                    
}}
\vskip .7cm

Before solving the Picard-Fuchs equations, one must choose a basis that 
depends both on the open as well as closed string phase.  A priori
there are many inequivalent basis choices, but in this model the
requirement that the open string expansion converge in a particular 
phase limits the choices.  Requiring the
expansions in the $z_i$ to converge in a neighborhood of the origin yields a 
unique solution for the particular phase.  By uniqueness the solution solves
the equations in any basis obtained from the original basis by a linear
transformation in which the transformation matrix has no negative
entries.  The
solution may not be unique in any particular basis.  Also, even when the
expansions are convergent in a particular basis, there may be a 
correction which is ``nonperturbative'' with respect to that basis and 
necessary for integrality.  We have found a unique basis for this phase 
in which the above problem does not occur.  We need to define $r_9=r_6-r_7$
for this basis.
The solutions are as follows:
\eqn\thitwo{\eqalign{z_1&={q_1 e^{-N+T}\over 1+q_1 q_3}\cr 
z_2&=q_2(1+q_1 q_3)e^{-N+M-T}\cr z_3&={q_3 e^{N-T}\over 1+q_1 q_3}\cr
z_7&=q_7(1+q_2 q_3 q_9) e^{-P+M-T}\cr z_8&=q_8 e^{N+P-2M}\cr
z_9&={q_9 e^{-M+2T}\over  (1+q_2 q_3 q_9)^2}\cr
\Delta&=(t_1+t_2+2t_7+2t_8+t_9)^2+{\rm instanton\
corrections}\cr}}
where $\Delta$ is a solution corresponding to a four-cycle, 
$$T=\sum_{m,n,p,q,r,s}{(-1)^{n+p+r+m}(n+m+p-r-2q-1)!z_1^r z_2^p z_3^m 
z_7^n z_8^s z_9^q\over m! q! r! 
(m-p-r+s)!(s-n)!(p-m-r)!(n-2q)!(p-2s+n-q)!}\, ,$$
$$N=\sum_{r,n,s}{(-1)^{s}(2r-s-1)!z_1^r z_2^r z_7^{2n} 
z_8^s z_9^{n}\over n! r!(s-2n)!(r+n-2s)!}\, ,$$
$$M=\sum_{r,n,s}{(-1)^{n+r}(2s-n-r-1)!z_1^r z_2^r z_7^{2n} z_8^s z_9^n\over
r! n!(s-2r)!(s-2n)!}\, ,$$ and 
$$P=\sum_{r,n,s}{(-1)^{s}(2n-s-1)!z_1^r z_2^r z_7^{2n}
z_8^s z_9^{n}\over n! r!(s-2r)!(r+n-2s)!}$$ and the $z_i$'s must be found as a
function of the $q_i$'s perturbatively. 

Examining Diagram (5) one sees that there is a reflection symmetry
about a line through the equator of ${\bf P}^1 (r_4)$ and ${\bf P}^1 (r_8)$.
The above solutions do not reflect this symmetry because any choice of
basis necessarily breaks this symmetry.  In this case the selection of
$q_3$ breaks the symmetry.  We had previously chosen the basis with 
$q_6$ instead of $q_9$ and found that the open string expansion did
not give integers without the term $(1+q_2 q_3 q_9)=(1+q_2 q_3 q_6/q_7)$,
but the perturbative solution, ${\hat t}_6$, of the Picard-Fuchs equations  
does not converge if we include this term.  The pieces of the solution 
involving negative powers solve the Picard-Fuchs equations by themselves
so there are
ambiguites of the solution in general.  Clearly, the
``nonperturbative'' pieces are essential for integrality of the numerical
invariants.
We also note that this model is the only one
treated so far where the above ambiguity involving negative powers
occurs.  In different phases of the theory we need to resolve the Picard-Fuchs
equations.  

Up to total order fourteen in the $q_i$'s (linear order in $q_9$)
we find that the coefficients in
the expansions of the inverse mirror map are integral.
The expansions in this phase are ($z_1 z_3={q_1 q_3\over (1+q_1 q_3)^2}$)

\eqn\fourexpanse{\eqalign{z_1&=q_1+q_1 q_2 q_3+
q_1 q_2 q_3 q_8+q_1 q_2 q_7 q_8-2q_1q_2 q_3 q_7 q_8
+\cdots\cr
z_2&=q_2+q_1 q_2 q_3+q_2 q_8+q_1 q_2 q_3 q_8+
+\cdots\cr z_7&=q_7-q_2 q_3 q_7+q_7 q_8+q_1 q_2 q_7 q_8-2
q_2 q_3 q_7 q_8+q_2 q_3 q_7 q_9+q_2 q_3 q_7 q_8 q_9+
\cdots\cr z_8&=q_8+q_1 q_2 q_8+
\cdots\cr z_9&=q_9-2 q_2 q_3 q_9-q_8 q_9+q_1 q_2 q_8 q_9-2
q_2 q_7 q_8 q_9+4q_2 q_3 q_7 q_8 q_9-2q_1 q_2 q_3 q_7 q_8 q_9+
\cdots\cr}.}

In the limit that $z_7=z_8=z_9=0$, $e^N=1+q_1 q_2$, $e^T=1+q_2 q_3$,
and the solutions are precisely those of the ${\bf Z}_2\times {\bf Z}_2$
case.  The limit $z_1=z_2=z_3^{-1}=z_7=z_8=z_9=0$ while $z_2 z_3 z_7 z_8$
is finite yields the solution for the blowup of the ${\bf Z}_3$
orbifold.  Taking $z_1=z_2^{-1}=z_3=z_7^{-1}=z_8=z_9=0$ with $z_2 z_3$ and
$z_2 z_7 z_8$ finite yields
the ${\bf P}^1\times {\bf P}^1$ case.  

\subsec{The Mirror and Open String Calculations}

Putting $y_2=u$, $y_6=v$, and $y_7=0$ the equation for the mirror is
\eqn\thithree{\eqalign{xz&=P(u,v)=1+e^u+e^v+e^{u+v-t_1}+e^{u-v-t_2}
 +e^{v-u-t_3}+\cr &\qquad
+e^{-v-t_2-t_7-t_8}+e^{-2v+u-2t_2-t_8}+e^{-3v+u-3t_2-2t_7-2t_8-t_9}\cr}.}
This choice is sensible for open string phases on the four-cycle.
Notice again that this equation reproduces the ${\bf Z}_2\times {\bf Z}_2$
case when $z_7=z_8=z_9=0$.  To obtain the standard version of the
${\bf Z}_3$ case, take $v'=v-u-t_3$
and the previously discussed limit.  The standard version of 
${\bf P}^1\times {\bf P}^1$ results from $u'=u-v-t_2$ and the above limit.
The Riemann surface $P(u,v)=0$ 
can be visualized
in the following diagram.

\bigskip
\centerline{\epsfxsize=0.5\hsize\epsfbox{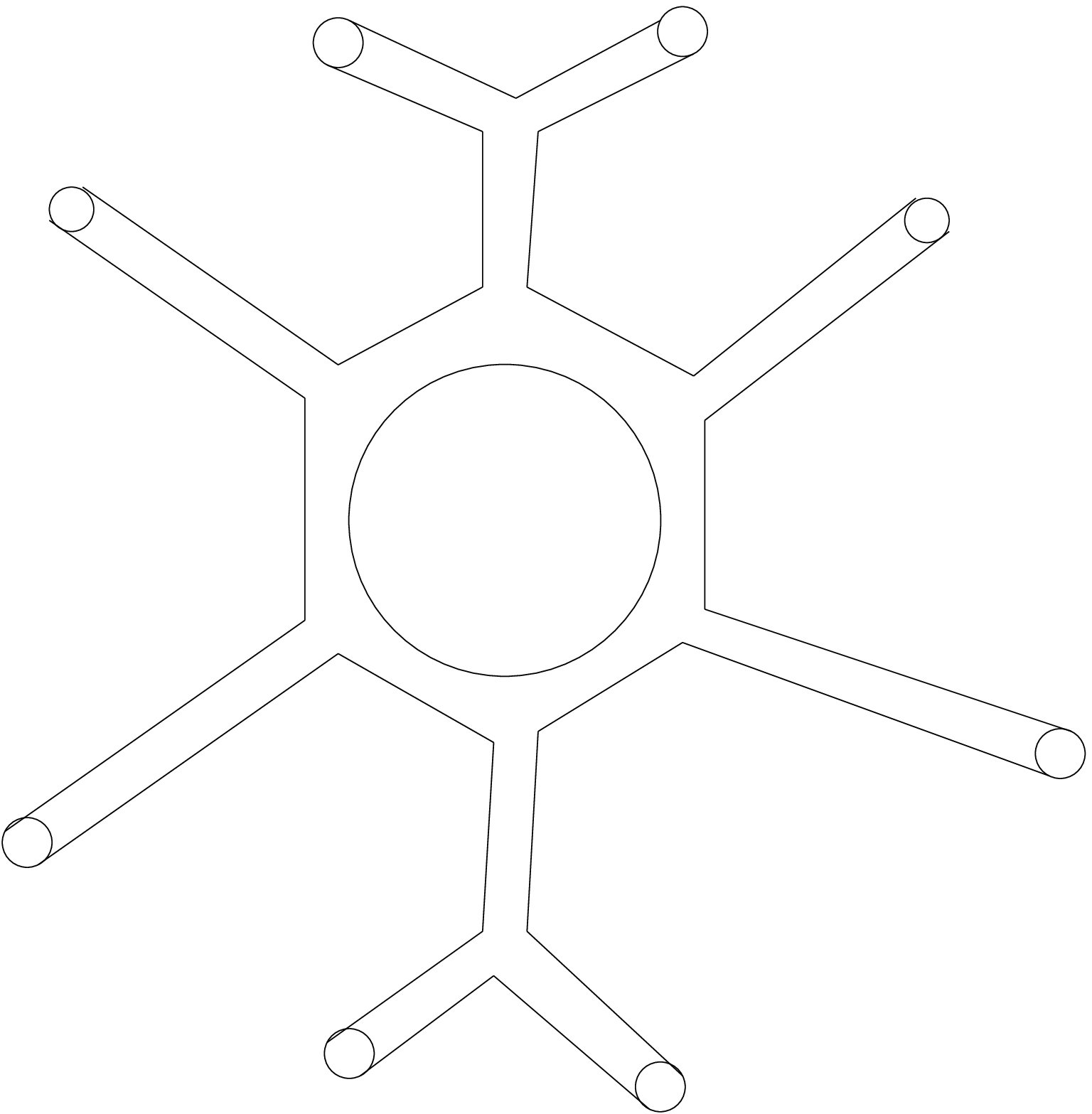}}
\bigskip
\centerline{\vbox{\noindent{\bf Diag. 8.}
Riemann Surface $P(u,v)=0$ in mirror of ${\bf Z}_2\times{\bf Z}_4$ Blowup
}}
\vskip .7cm

A noncompact, supersymmetric three-cycle 
intersecting
the toric base of the original manifold is determined by $|x_2|^2-|x_7|^2=c_1$,
$|x_6|^2-|x_7|^2=c_2$, and $\sum_i Arg(x_i)=0$.  There are sixteen phases
for the Lagrangian three-cycle with symmetries relating the phases in the 
lower half of diagram nine to those in the upper half.  For instance the phase 
with a three-cycle intersecting ${\bf P}^1(r_7)$ is equivalent to phase I 
under the obvious 
permutation of the $r_i$'s and $v\rightarrow -v$.  Also, phases II and III
are equivalent.  There are, accordingly,
nine inequivalent phases.

\centerline{\epsfxsize=0.5\hsize\epsfbox{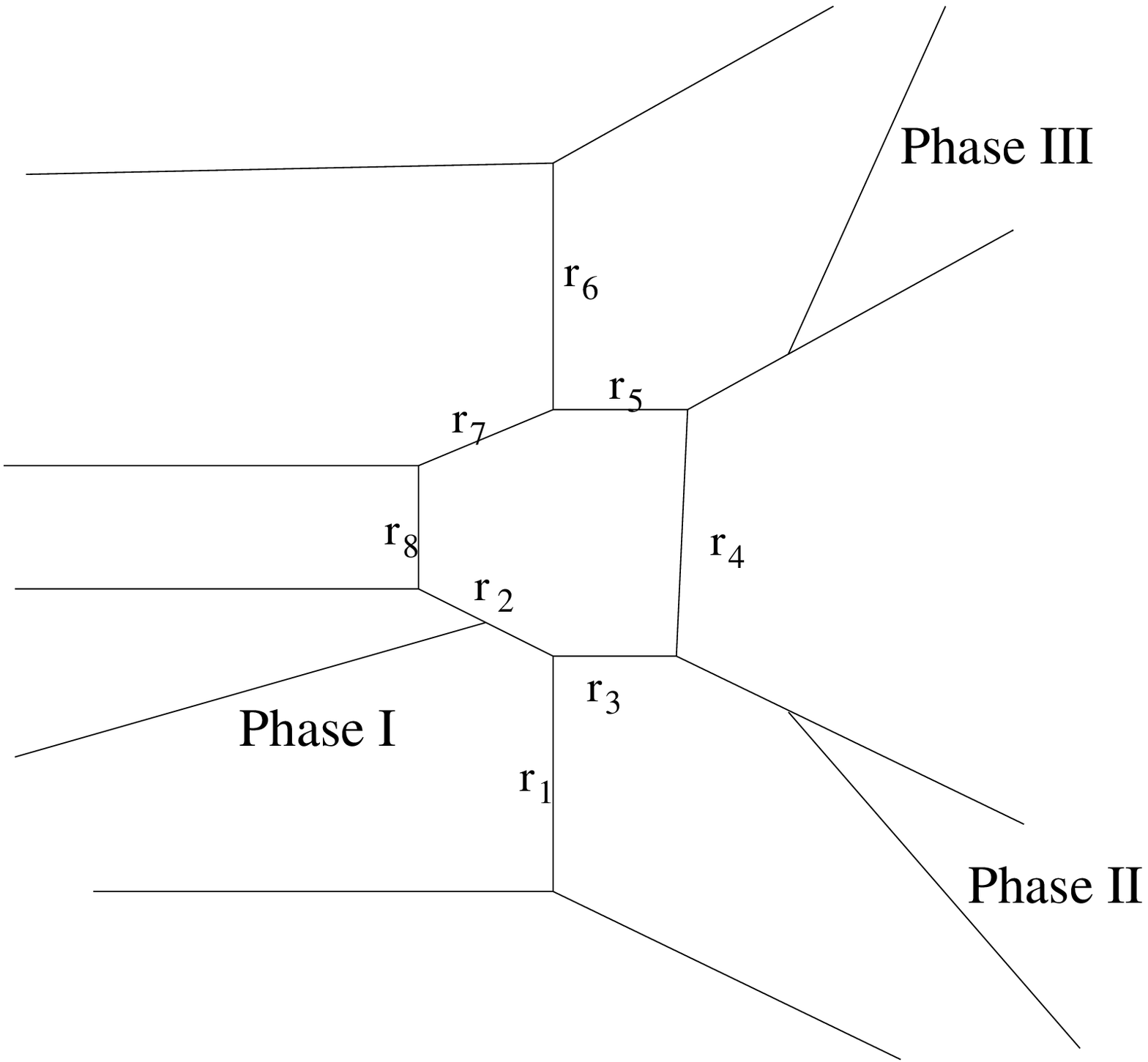}}
\bigskip
\centerline{\vbox{\noindent{\bf Diag. 9.}
Phases for noncompact three-cycle in ${\bf Z}_2\times{\bf Z}_4$ Blowup
}}
\vskip .7cm

Five of these phases can be parametrized by $u=0$ after a change of coordinates
while the rest correspond to $v=0$.  The latter yield quartic equations which
can be solved and expanded, but the amount of computer time required seems
prohibitively large.  Fortunately, the $u=0$ phases are quadratic.  These
are still quite complicated because the expansion involves seven variables.
We will, thus, restrict ourselves to two of the $u$ phases and only test the
integer hypothesis at low order in the expansion.  Phase I corresponds to
$c_1=0$ and $0<c_2<r_2$.  In phase II  
$c_1=-r_3$ and $c_2<<0$.  In phase I we calculate the 
zero mode piece of $u$ to be 
\eqn\thifour{\delta u=i\pi +
{t_1-{\hat t}_1\over 2}-{t_3-{\hat t}_3\over 2}=i\pi -T+N .}
We can determine $\delta v$ by expanding an adjacent phase in which $u-v=0$,
and we find 
\eqn\thifive{\eqalign{\delta v &=i\pi+{t_1-{\hat t}_1\over 4}-
{3(t_2-{\hat t}_2)\over 4}-{t_7-{\hat t}_7\over 2}-
{t_8-{\hat t}_8\over 2}-
{t_9-{\hat t}_9\over 4}\cr &
\cr &=i\pi -T+\ln (1+q_1 q_3)\cr }.}
Solving \thithree\ for $u$ and substituting instanton corrected variables,
we can write the expansion for $\hat u$ in this phase.
\eqn\thisix{\eqalign{{\hat u}&=\sum\nolimits '_{n,p,q,r}[{(-1)^q (n-1)!
e^{-Nn} e^{Mr} e^{Pq}\over p!q! r!(n-p-q-r)!}\times\cr
\qquad &e^{{\hat v}(4p+2r+q-3n)} q_1^p q_2^{3n-3p-2r-q}   
q_7^{2(n-p-r-q)} q_8^{2n-2p-2r-q}q_9^{n-p-r-q} ]\cr
&-\sum\nolimits '_{a,b,c,d,e,m,n,p}[{(-1)^{b+d} (2n+m-1)!\over n!a!b!c!d!e!f!
(n-a-b-c-d)!(p-f)!(m-p-e)!}e^{-T(2n+m)}
e^{Mc} e^{Nb} e^{Pd}\times\cr
\qquad &(e^{{\hat v}(4a+3b+2c+d+m-2p-2n)}
q_1^{a+e}q_2^{3n-3a-3b-2c-d+p+f} q_3^{n+e+f}\times\cr
\qquad &  
q_7^{2(n-a-b-c-d)+p} q_8^{2n-2a-2b-2c-d+p} q_9^{n-a-b-c-d+f})]\cr}}

One can easily verify that
the above expansion reduces to \eleven\ in the limit $q_7=q_8=q_9=0$.  Also,
the limit $q_1=q_9=0$ yields the ${\bf F}_2$ blown up at two points.
One can also
show that the inner phase of the ${\bf Z}_3$ model with ambiguity
$n=1$ is achieved in the limit $q_1=q_2=q_3^{-1}=q_7=q_8=q_9=0$ with
$q_2 q_3 q_7 q_8$ finite and ${\hat v}\rightarrow{\hat v}+t_3$.  In choosing
this basis we have required that the expansion be convergent in a
neighborhood of the origin in both open and closed string variables.  In
phase I this requirement entails that negative winding terms of the
form $(\prod_i q_i^{n_i}) e^{-m{\hat v}}$ have $n_2\ge m$.  In an earlier
calculation we chose a basis not meeting this last requirement and
many terms had fractional invariants.  
Note that the basis with $q_6$ instead of $q_9$ does meet this latter 
requirement but still has fractions due to the nonperturbative piece.
Up to linear order in $q_9$, 
quadratic order in
$q_7$, cubic order in $q_1$ and $q_3$, quartic 
order in $q_2$
and sixth order in  $e^{\pm{\hat v}}$,
all of the numerical invariants in this phase are integral.
We present a table of terms corresponding to the homology classes that failed 
to be integral in the badly chosen basis.
\vfill\eject

\subsec {Numerical Invariants}

\centerline{\bf{Table 4}:}
\smallskip
\centerline{Numerical Invariants}
\smallskip
\begintable

$k$|$m_1$|$m_2$|$m_3$|$m_7$|$m_8$|$m_9$|$N_{k,{\vec m}}$
|$k$|$m_1$|$m_2$|$m_3$|$m_7$|$m_8$|$m_9$|$N_{k,{\vec m}}$\cr
$2$|$1$|$3$|$2$|$2$|$2$|$1$|$-64$|$-2$|$1$|$3$|$0$|$2$|
$2$|$1$|$0$\cr
$2$|$0$|$4$|$2$|$2$|$2$|$1$|$-36$|$-2$|$2$|$4$|$0$|$2$|$2$|$1$|
$0$\cr
$2$|$0$|$4$|$2$|$2$|$3$|$1$|$-36$|$-2$|$0$|$3$|$1$|$2$|$2$|$1$|
$1$\cr
$3$|$0$|$3$|$3$|$2$|$2$|$1$|$-621$|$-2$|$0$|$4$|$2$|$2$|$2$|$1$|
$4$\cr
$3$|$1$|$4$|$3$|$2$|$2$|$1$|$-2214$|$-3$|$0$|$3$|$0$|$2$|$2$|$1$|
$0$\cr
$3$|$1$|$4$|$3$|$2$|$3$|$1$|$-2214$|$-3$|$1$|$4$|$0$|$2$|$2$|$1$|
$0$\cr
$3$|$3$|$3$|$3$|$2$|$2$|$1$|$72$|$-3$|$0$|$3$|$0$|$2$|$2$|$1$|
$0$\cr
$4$|$1$|$3$|$2$|$2$|$2$|$1$|$-128$|$-3$|$1$|$4$|$0$|$2$|$2$|$1$|
$0$\cr
$4$|$3$|$3$|$2$|$2$|$2$|$1$|$0$|$-4$|$0$|$4$|$0$|$2$|$2$|$1$|
$0$\cr
$4$|$0$|$4$|$2$|$2$|$2$|$1$|$-56$|$-4$|$0$|$6$|$2$|$4$|$4$|$1$|
$0$\cr
$4$|$0$|$4$|$2$|$2$|$3$|$1$|$-56$|$-4$|$0$|$6$|$2$|$4$|$5$|$1$|
$0$\cr
$6$|$1$|$3$|$2$|$2$|$2$|$1$|$-224$|$6$|$3$|$3$|$3$|$2$|$2$|$1$|
$1026$\cr
$6$|$3$|$3$|$2$|$2$|$2$|$1$|$0$|$6$|$1$|$4$|$3$|$2$|$2$|$1$|
$-10656$\cr
$6$|$0$|$4$|$2$|$2$|$2$|$1$|$-84$|$6$|$0$|$4$|$2$|$2$|$3$|$1$|
$-84$\cr
$6$|$0$|$3$|$3$|$2$|$3$|$1$|$-936$|$6$|$1$|$4$|$3$|$2$|$3$|$1$|
$-10656$
\endtable

What is the meaning of these numerical invariants and why do we anticipate
that the numerical invariants of this phase are integers?  We consider the 
moduli space of maps from a Riemann surface of genus zero with one 
boundary into the Calabi-Yau such that the relative homology class
of the image is labeled by the wrapping number on each two-sphere of the
basis and the winding number around a noncontractible circle on the
Lagrangian three-cycle.  These maps should be holomorphic in the interior
of the disk.  The moduli space of these maps is generally noncompact, and
one must add in extra maps that may be singular to define a compact space.
There may be disconnected components of the moduli space when there are
homotopically inequivalent maps into some relative homology class.  One then
defines a cohomology class analogous to the Euler class, and the numerical 
invariant is obtained by integrating this class over the moduli space.
For the case of genus zero closed strings, one should fix three 
complex parameters corresponding to $SL(2,{\bf C})$ transformations 
of the complex plane while for disks one fixes three real parameters
corresponding to $SL(2,{\bf R})$ transformations of the upper half plane.
If the dimension of the moduli space is zero after this fixing, the maps
are isolated and the numerical invariants can be interpreted as 
counting curves.  Otherwise, the integral over the moduli space could give
fractions when there are orbifold singularities in the moduli space.
Of course, there are many technicalities needed to make the above
discussion rigorous.

A first principles calculation from the nonlinear sigma model point of view
as described above is generally difficult.  In this paper our determination of
the invariants has been facilitated by an equivalent calculation on the
local mirror.  The drawback is that one does not have a direct argument that 
the invariants should be integers.  Another approach is to start with the 
boundary linear sigma model that flows to the conformally invariant 
nonlinear sigma model at low energies.  The relevant correlation
functions that yield the numerical invariants are in a topological sector
of the theory.  The corresponding correlators in a topologically
twisted version of the linear sigma model are scale invariant so the
two calculations should be equivalent.  The correlators are intersection
forms on the moduli space of classical solutions with a given instanton
number $m_i={1\over 2\pi}\int_{\Sigma}F_i$ and winding number $k_p=
{1\over 2\pi}\int_{\partial\Sigma}\Lambda_p$ where $F_i$ is the gauge field 
strength which is integrated over the two-dimensional worldsheet 
$\Sigma$ with boundary $\partial\Sigma$ and $\Lambda_p$ 
is a boundary field which
couples to the theta angle.  It would be interesting to calculate
correlators in the boundary linear sigma model corresponding to the numerical
invariants.  This correspondence between the two theories has been
verified in several examples for closed worldsheets.  Our aim here is to
discuss the structure of the boundary linear sigma model moduli space
in order to determine a criterion for integrality of the intersection
forms.  Assuming the correspondence is valid, we can apply this criterion
to the nonlinear sigma model.  Note that the moduli spaces of the two
theories are different, but the correlators of this topological sector 
are expected to be the same.  

In the linear sigma model the moduli space can be described as a space of 
holomorphic sections $x_i$ of a line bundle over the upper half complex plane 
$H$ of degree $d^i=\sum_j(m_j l_j^i+k_p l_p^{(1)i})$ (cf. \laterone).  
One usually calls
this bundle ${\cal O}(d^i)$.  If $d^i\ge 0$,  one can write a section as
$x_i=\sum_{n=0}^{d^i}x_{in} z^n$ where $z$ is a coordinate on $H$ and the
$x_{in}$ must be chosen so that solutions of $x_i=0$ all lie in $H$.  If
$d^i<0$, there is no holomorphic section and $x_i=0$.  Note that we
have tried to indicate ${\cal O}(0)$ deformations of two-spheres
on the toric diagram 
by drawing adjacent lines parallel.  The moduli space
takes the form 
\eqn\thieight{M_{{\vec m},{\vec k}}={(X_{{\vec m},{\vec k}}
-I_{{\vec m},{\vec k}})\over G}}
where $X_{{\vec m},{\vec k}}$ is the space of $x_{in}$ such that 
$x_i$ has zeroes in $H$.  The maps $x_i$ are singular whenever for some 
$z$, the values of $x_i$ are ``impossible''.  We can readily see what is 
meant by ``impossible'' by examining Diagram (5).  If we denote 
$D_i=\{(x_1,x_2,\cdots,x_9)|x_i=0\}$, then $I=\{ (x_1,x_2,\cdots,x_9)\in
\cap_{l=1}^k D_{l}|\cap_{l=1}^k D_{l}\cap M=\phi\}\cup O$ where $M$ is
the toric manifold corresponding to Diagram (5).  Then $I_{{\vec m},{\vec k}}$
is the set of $x_{in}$ such that for all $z$, $(x_1,x_2,\cdots)\in I$.
The moduli space is compactified by including maps where some points
on the worldsheet are mapped into $I$.  In general the moduli space
could still be noncompact, but noncompact ${\cal O}(0)$ directions are
generally cut off by the disk.  For ${\vec k}=0$, the moduli space is
the Lagrangian three-cycle.
Here $O$ is determined by the disk constraints.  For instance, in phase I
of our case , $O=\{(x_1,x_2,\cdots,x_9)|D_3\cup D_6\cup (D_7\cap(M-D_2))
\cup (D_2\cap(M-D_7))\}$.
We mod out the set of allowed $x_{in}$ by complex gauge transformations
$G:x_{im}\rightarrow\prod_{j,p}(\alpha_j^{l_j^i}\beta_p^{l_p^{(1)i}}
\gamma^{l^{(2)i}})x_{im}$ where $\alpha_j\in {\bf C}^*$, $\beta_p\in 
{\bf R}^+$,
and $\gamma\in U(1)$(cf. \laterone\latertwo).  
Writing $x_i=r_0(z-r_1)\cdots(z-r_n)$, we see
that this makes sense because we require $r_1,\cdots,r_n\in H$ but 
$r_0\in {\bf C}$ and $r_0\ne 0$ if $x_i\not\equiv 0$.  The rescalings and
rotations
preserve the zeroes of $x_i$, and the condition $\sum_i Arg(x_i)=0$
is redundant at the boundary of the disk on the toric base so $\gamma\equiv 1$
for disks stuck at the base.  Note that all disks in the geometric phase
are stuck at the base.  We have shown that the moduli space is well defined.
It is now easy to see that $I$ includes any regions that are fixed by
$G$ unless we shrink some of the two-spheres to enter a nongeometric,
orbifold phase.  Since the moduli space of the geometric phase is
smooth, the intersection form must give integers.  On the other hand, the
moduli space is frequently singular in nongeometric phases and fractions are
possible.  One may expect fractional terms when the boundary of the disk
is fixed by the orbifold. 

In phase II our coordinate transformation is 
$u\rightarrow u'=u-t_3$, $v\rightarrow v'=v$.  We obtain 
$\delta u=i\pi+{\delta t_1\over 2}+{\delta t_3\over 2}=i\pi +
\ln (1+q_3)$  and $\delta v$ as in \thifive .  This
phase reproduces precisely the outer phase (the three cycle intersects a
noncompact two-cycle) of the ${\bf Z}_3$ case in the
${\bf Z}_2$ reflection of the limit discussed previously.  
This phase also reduces to the outer phase of the blown up ${\bf F}_2$ in
the limit that $q_1=q_9=0$.
All of the corresponding terms are integral.  In this phase $v'<0$ and
all terms have $k<0$.  Calculating to the same order
as in phase I, we find that all terms are integral.

We have also calculated disk instantons in the nongeometric phase 
obtained by flopping ${\bf P}^1 (r_6)$ and then flopping ${\bf P}^1 (r_6+r_7)$
at $x_4=x_7=0$.  In this phase $x_4>0$ so the flopped ${\bf P}^1$ is
nongeometric.  The calculation gives half-integer invariants for terms of the 
form  $q_2^4 q_3^2 q'_8 e^{2n{\hat v}}$.  
($r_8'=r_8-r_2+r_4+r_6$, $n\in {\bf Z}^+$)  These results 
are puzzling.  Although there is a ${\bf Z}_2$ orbifold
singularity on the ``geometric'' ${\bf P}^1 (r_8')$ in this phase, the
presence of the disk prevents the moduli space from being an
orbifold.  We believe the resolution of this paradox is that there is
a ``nonperturbative'' contribution corresponding to $P$ \thitwo\ .
The relative homology classes in question are present in Table 4.  
Depending on the choice of coordinates, one can have square root
branch cuts ($\sqrt{q_8'}$) in the geometric phase but one can always
find coordinates without branch cuts.  The correction $P$ is an expansion
in the flopped Kahler parameter for which we need the analytic continuation
to this phase.  There are other terms in this phase in which the moduli space
does contain orbifold singularities ($x_3=x_5=0$, $x_2\not\equiv 0$),
and we anticipate fractional invariants.  
We have not pursued this calculation further.

We have in this model a flop transition that augments the four-cycle from an
${\bf F}^2$ blown up at two points to one blown up at four points. 

\centerline{\epsfxsize=0.5\hsize\epsfbox{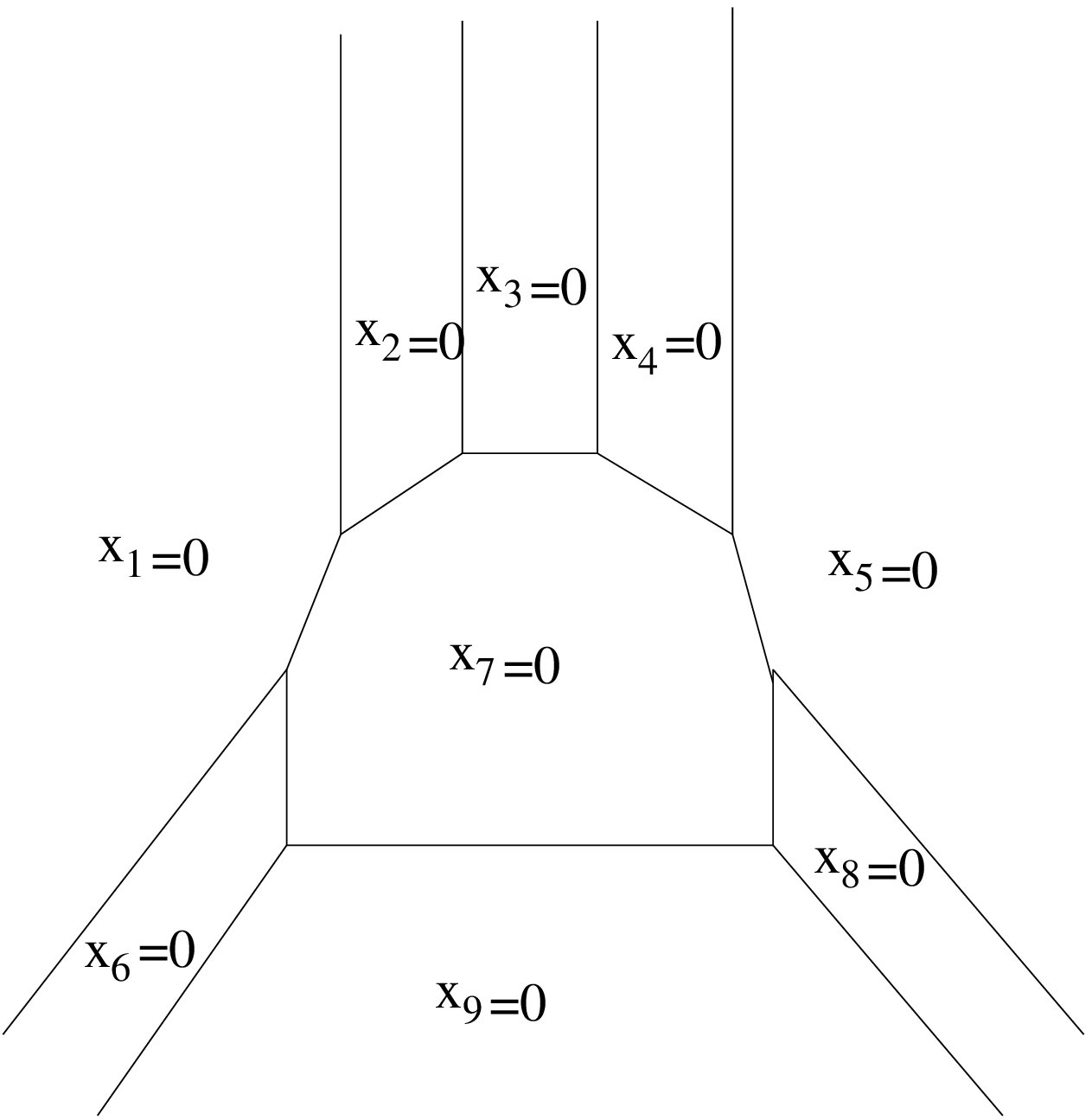}}
\bigskip
\centerline{\vbox{\noindent{\bf Diag. 10.}
Flopped Phase of ${\bf Z}_2\times{\bf Z}_4$ Blowup:$r_1\rightarrow -r_1$,
$r_6\rightarrow -r_6$}}
\vskip .7cm

We
can flop $r_1$ and $r_6$ to obtain this phase.  The conifold
type transitions occur for the part of the diagram that reduces 
to the ${\bf Z}_2\times {\bf Z}_2$ case.  As in that case one can substitute
${\bf S}^3$'s for ${\bf P}^1$'s.  Our amplitudes ${\cal F}_{0,1}$ should
correspond to Chern-Simons theory in a complicated background.

\newsec{${\bf Z}_7$}

The ${\bf Z}_7$ case is interesting because there are three adjacent 
four-cycles.  However, there are no geometric flops unlike the previous
examples.  The charges of six chiral fields under three $U(1)$'s are
\eqn\foone{\eqalign{l_1&=(0,1,-3,0,1,1)\cr l_2&=(1,-2,1,0,0,0)\cr
l_3&=(-2,1,0,1,0,0).\cr}}
Solving the equations $\sum_j l_i^j |x_j|^2=r_i$ generates the following
toric diagram where $r_4=r_1+3 r_2$ and $r_5=r_1+3 r_2+5 r_3$.

\centerline{\epsfxsize=0.5\hsize\epsfbox{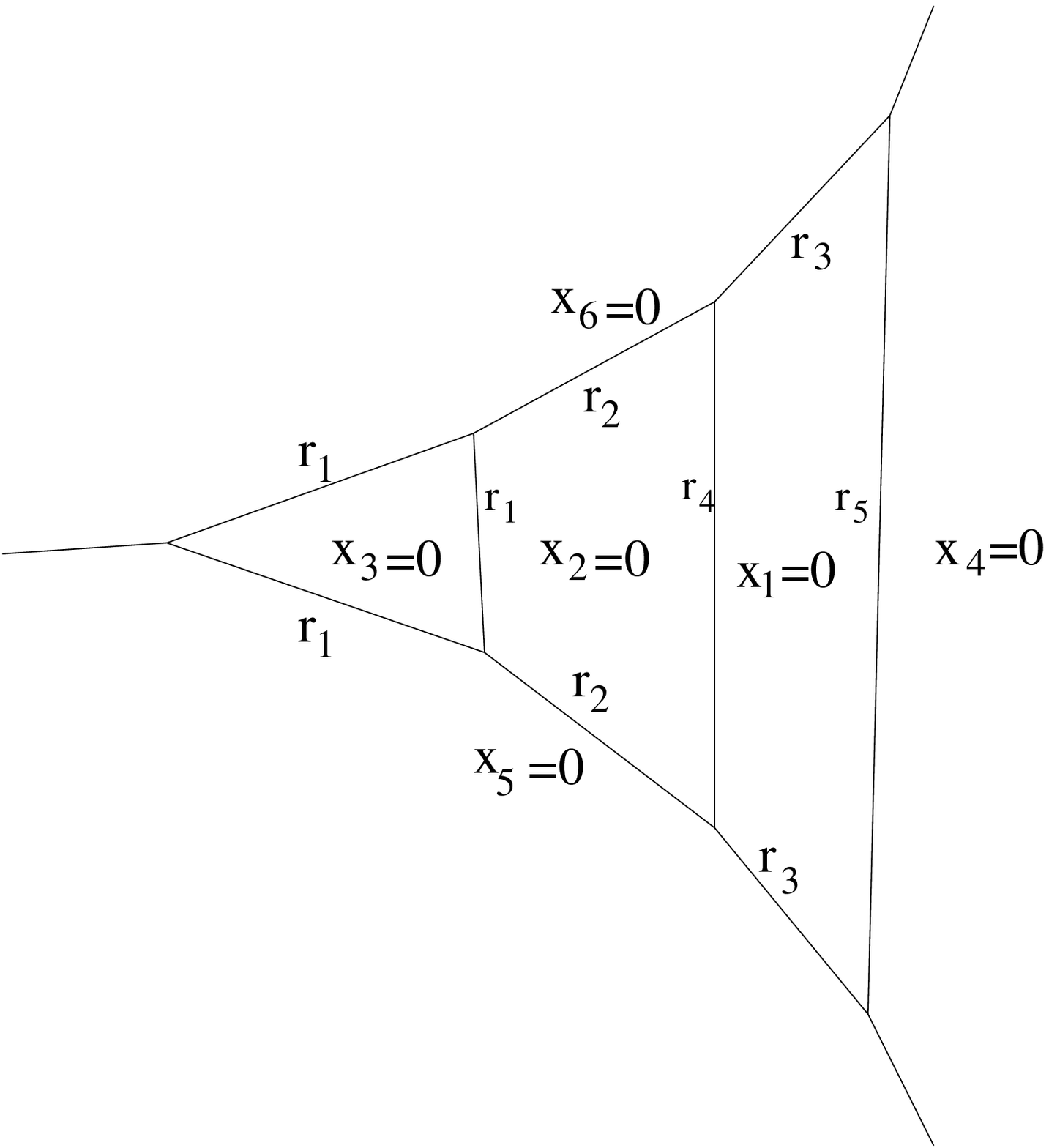}}
\bigskip
\centerline{\vbox{\noindent{\bf Diag. 11.}
Toric Diagram of ${\bf Z}_7$ Blowup
}}
\vskip .7cm

The instanton corrected
Kahler parameters are given by
\eqn\fotwo{\eqalign{z_1&=q_1 e^{-3M+P}\cr z_2&=q_2 e^{M-2P+Q}\cr
z_3&=q_3 e^{-2Q+P}\cr}}
where
 
$$M=\sum_{n,m,p}{(-1)^{n+m}(3n-m-1)!z_1^n z_2^m z_3^p\over n!^2 p!
(m-2p)! (n-2m+p)!}\, ,$$

$$P=\sum_{n,m,p}{(-1)^{m+p}(2n-m-p-1)!z_1^p z_2^n z_3^m\over
p!^2 (n-3p)! m! (n-2m)!}\, , $$
{\rm and}
$$Q=\sum_{n,m,p}{(-1)^m(2n-m-1)!z_1^p z_2^m z_3^n\over n! p!^2 (m-3p)!
(n-2m+p)!}.$$

  These expressions for $q_i$ must be inverted to
obtain $z_i(q_i)$.  The expansions are
\eqn\expaseven{\eqalign{z_1&=q_1+6q_1^2+q_1 q_2+10 q_1^2 q_2+4q_1^2 q_2^2+
q_1 q_2 q_3+10 q_1^2 q_2 q_3+8 q_1^2 q_2^2 q_3+4q_1^2 q_2^2q_3^2+\cdots\cr
z_2&=q_2-2 q_1 q_2+5q_1^2 q_2-2q_2^2+6q_1q_2^2-20q_1^2 q_2^2+q_2 q_3
-2q_1 q_2q_3+5q_1^2q_2 q_3\cr &\qquad-3q_2^2q_3+8q_1q_2^2q_3-26q_1^2 q_2^2 q_3
-2q_2^2 q_3^2+6q_1 q_2^2 q_3^2-20q_1^2q_2^2 q_3^2+\cdots\cr
z_3&=q_3+q_2 q_3-2 q_1 q_2q_3+5q_1^2 q_2q_3-2q_1 q_2^2 q_3 -2q_3^2\cr &\qquad-
3q_2q_3^2+6q_1 q_2 q_3^2-15q_1^2 q_2q_3^2-2q_2^2 q_3^2+12q_1 q_2^2 q_3^2-44
q_1^2q_2^2 q_3^2+\cdots\cr}}

There are also three more solutions to the Picard-Fuchs
equations corresponding to four-cycles.  The ${\bf Z}_3$ and ${\bf Z}_5$
cases can be obtained in the appropriate limit.  To obtain ${\bf Z}_3$
we set $z_2=z_3=0$.

The equation for the mirror can be written as
\eqn\fothree{xz=P(u,v)=1+e^u+e^v+e^{-u-v-t_1}+e^{2u-t_2}+e^{3u-2t_2-t_3}}
where $y_2=u$, $y_5=v$, and $y_3=0$. The Riemann surface that describes
moduli of the three-cycle is a genus three surface with legs extending to
infinity.  

\centerline{\epsfxsize=0.5\hsize\epsfbox{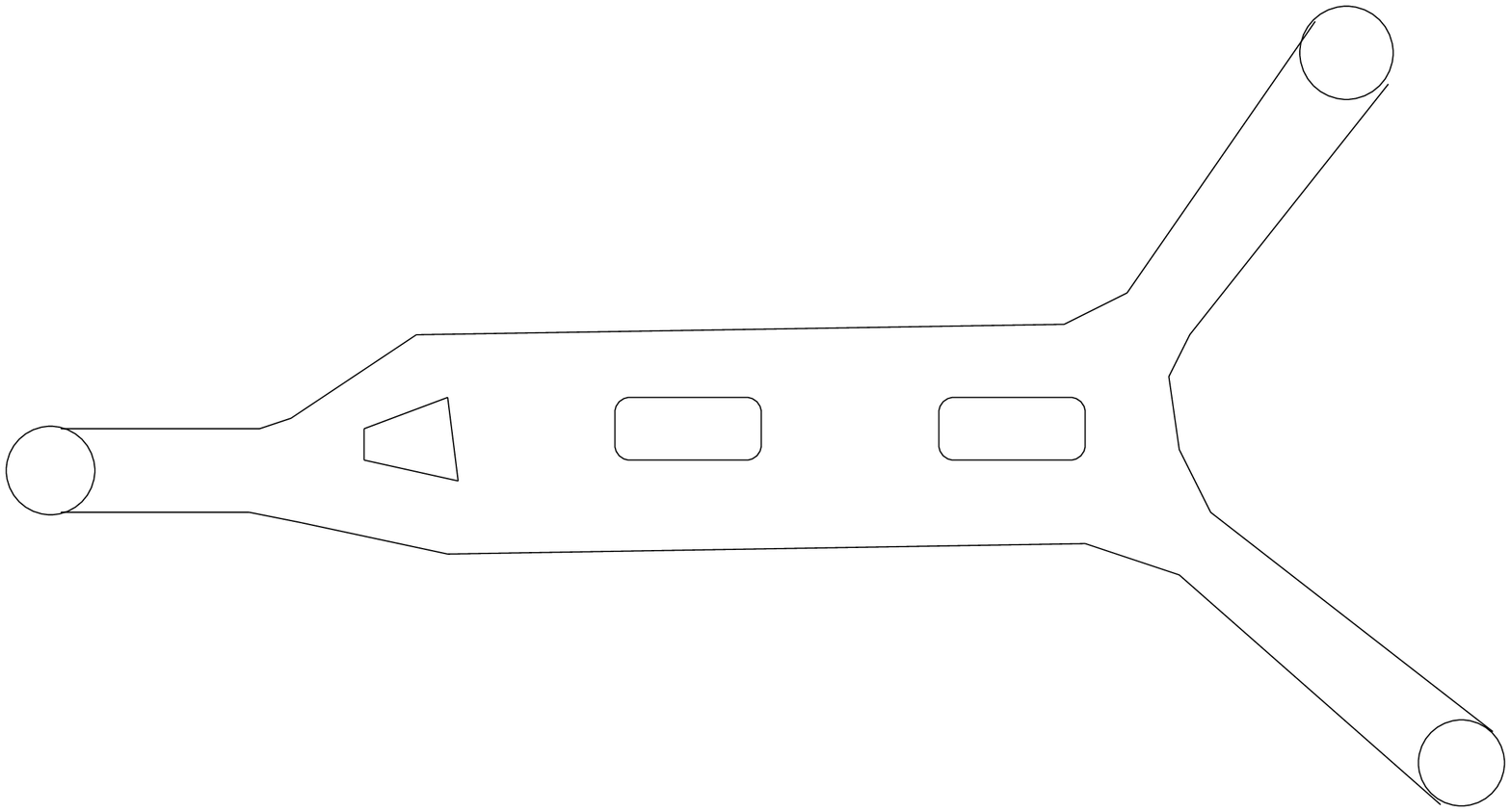}}
\bigskip
\centerline{\vbox{\noindent{\bf Diag. 12.}
Riemann Surface $P(u,v)=0$ in mirror of ${\bf Z}_7$ Blowup
}}
\vskip .7cm

\noindent Clearly, we recover the ${\cal O}(-3)
\rightarrow {\bf P}^2$ (${\bf Z}_3$ blowup) when $z_2=z_3=0$.  Letting our Lagrangian
three-cycle be determined by $|x_2|^2-|x_3|^2=c_1$, 
$|x_5|^2-|x_3|^2=c_2$, and $\sum_i Arg(x_i)=0$; we find eight inequivalent 
phases taking into account the obviously symmetric phases. 

\centerline{\epsfxsize=0.5\hsize\epsfbox{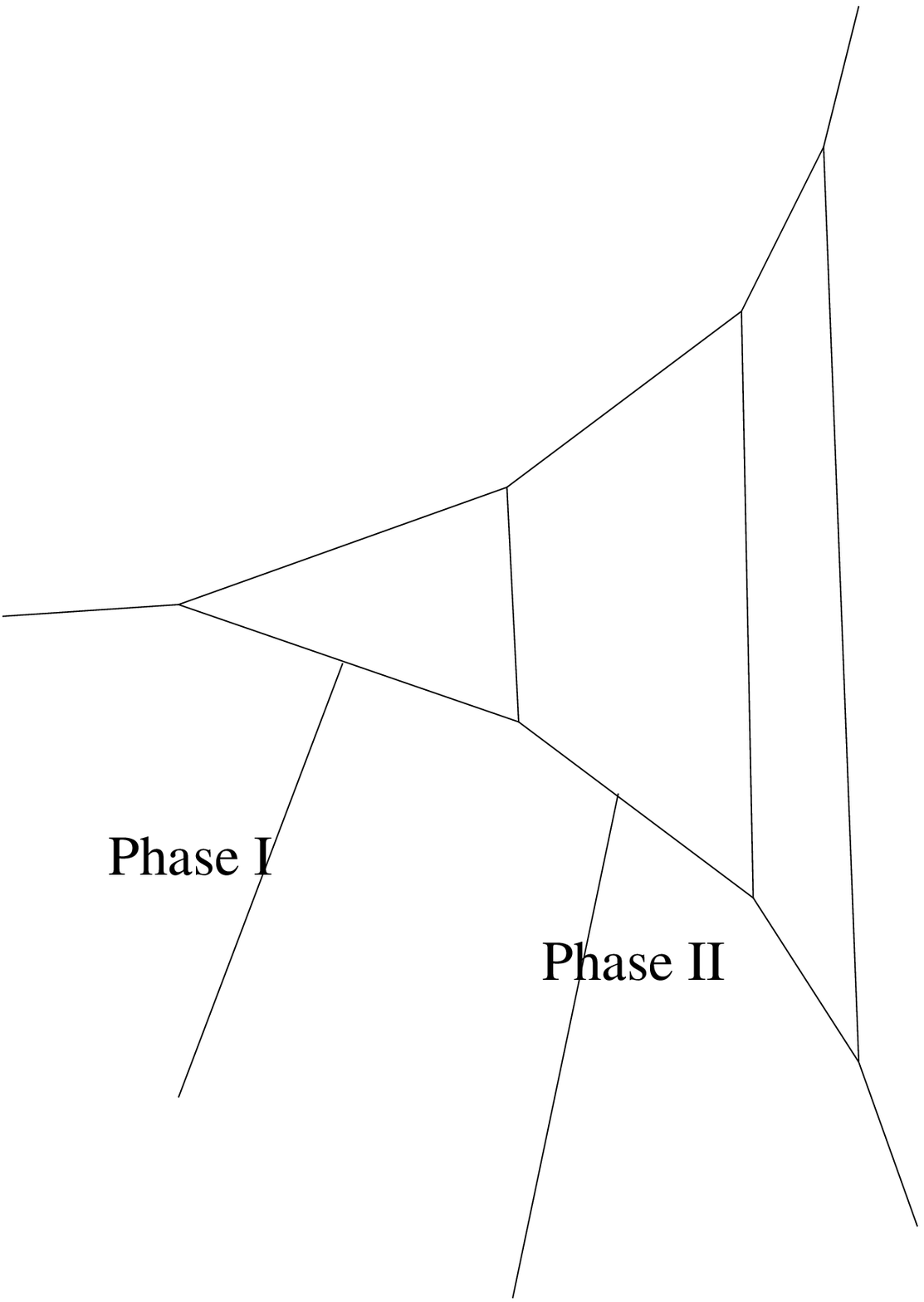}}
\bigskip
\centerline{\vbox{\noindent{\bf Diag. 13.}
Phases for noncompact three-cycle in ${\bf Z}_7$ Blowup
}}
\vskip .7cm 

\noindent Again some of the phases are described by 
quadratic equations, whereas
others require a quartic equation.  Phase I has $c_2=0$ and
$c_1\approx {r_1\over 2}$ while in phase II, $c_1=c_2\approx {-r_2\over 2}$.
In phase I
\eqn\fofour{\delta u=i\pi +{1\over 7}[-\delta t_1+4\delta t_2+2\delta t_3 ]=
i\pi -M+P}
and
\eqn\fofive{\delta v=i\pi +{1\over 7}[-3\delta t_1-2\delta t_2-\delta t_3 ]=
i\pi -M .}
In phase II $\delta u$ is unchanged while $\delta v'=\delta v-\delta u=-P$.
The results are shown in the tables.
\vfill\eject

\centerline{\bf{Table 5}:}
\smallskip
\centerline{Phase I:$m_1=2$,\ $k=3$}
\smallskip
\begintable

$m_3$| $m_2$=$0$|$1$|$2$|$3$|$4$\cr
$0$|$-9$|$-26$|$-49$|$-76$|$-110$\cr
$1$|$0$|$-26$|$-76$|$-144$|$-232$\cr
$2$|$0$|$0$|$-49$|$-144$|$-284$\cr
$3$|$0$|$0$|$0$|$-76$|$-236$\cr
$4$|$0$|$0$|$0$|$0$|$-118$

\endtable

\centerline{\bf{Table 6}:}
\smallskip
\centerline{Phase I:$m_1=2$,\ $k=4$}
\smallskip
\begintable

$m_3$| $m_2$=$0$|$1$|$2$|$3$|$4$\cr
$0$|$-12$|$-36$|$-72$|$-116$|$-172$\cr
$1$|$0$|$-36$|$-112$|$-224$|$-372$\cr
$2$|$0$|$0$|$-72$|$-224$|$-460$\cr
$3$|$0$|$0$|$0$|$-116$|$-376$\cr
$4$|$0$|$0$|$0$|$0$|$-180$

\endtable

\centerline{\bf{Table 7}:}
\smallskip
\centerline{Phase I:$m_1=2$,\ $k=5$}
\smallskip
\begintable

$m_3$| $m_2$=$0$|$1$|$2$|$3$|$4$\cr
$0$|$-15$|$-48$|$-99$|$-166$|$-250$\cr
$1$|$0$|$-48$|$-156$|$-324$|$-552$\cr
$2$|$0$|$0$|$-99$|$-324$|$-684$\cr
$3$|$0$|$0$|$0$|$-166$|$-556$\cr
$4$|$0$|$0$|$0$|$0$|$-258$

\endtable

\vfill\eject
\centerline{\bf{Table 8}:}
\smallskip
\centerline{Phase I:$m_1=2$,\ $k=6$}
\smallskip
\begintable

$m_3$| $m_2$=$0$|$1$|$2$|$3$|$4$\cr
$0$|$-19$|$-62$|$-132$|$-226$|$-349$\cr
$1$|$0$|$-62$|$-208$|$-444$|$-772$\cr
$2$|$0$|$0$|$-132$|$-444$|$-960$\cr
$3$|$0$|$0$|$0$|$-226$|$-776$\cr
$4$|$0$|$0$|$0$|$0$|$-355$

\endtable

\centerline{\bf{Table 9}:}
\smallskip
\centerline{Phase II:$m_1=2$,\ $k=-3$}
\smallskip
\begintable

$m_3$| $m_2$=$0$|$1$|$2$|$3$|$4$\cr
$0$|$-3$|$-8$|$-19$|$-46$|$-110$\cr
$1$|$0$|$-8$|$-28$|$-84$|$-232$\cr
$2$|$0$|$0$|$-19$|$-84$|$-284$\cr
$3$|$0$|$0$|$0$|$-46$|$-236$\cr
$4$|$0$|$0$|$0$|$0$|$-118$

\endtable

\centerline{\bf{Table 10}:}
\smallskip
\centerline{Phase II:$m_1=2$,\ $k=-4$}
\smallskip
\begintable

$m_3$| $m_2$=$0$|$1$|$2$|$3$|$4$\cr
$0$|$-4$|$-12$|$-32$|$-76$|$-172$\cr
$1$|$0$|$-12$|$-48$|$-144$|$-372$\cr
$2$|$0$|$0$|$-32$|$-144$|$-460$\cr
$3$|$0$|$0$|$0$|$-76$|$-376$\cr
$4$|$0$|$0$|$0$|$0$|$-180$

\endtable

\vfill\eject

\centerline{\bf{Table 11}:}
\smallskip
\centerline{Phase II:$m_1=2$,\ $k=-5$}
\smallskip
\begintable

$m_3$|$m_2$=$0$|$1$|$2$|$3$|$4$\cr
$0$|$-5$|$-18$|$-49$|$-116$|$-250$\cr
$1$|$0$|$-18$|$-76$|$-224$|$-552$\cr
$2$|$0$|$0$|$-49$|$-224$|$-684$\cr
$3$|$0$|$0$|$0$|$-116$|$-684$\cr
$4$|$0$|$0$|$0$|$0$|$-258$

\endtable

\centerline{\bf{Table 12}:}
\smallskip
\centerline{Phase II:$m_1=2$,\ $k=-6$}
\smallskip
\begintable

$m_3$|$m_2$=$0$|$1$|$2$|$3$|$4$\cr
$0$|$-7$|$-26$|$-36$|$-166$|$-347$\cr
$1$|$0$|$-26$|$-112$|$-324$|$-772$\cr
$2$|$0$|$0$|$-36$|$-324$|$-960$\cr
$3$|$0$|$0$|$0$|$-166$|$-776$\cr
$4$|$0$|$0$|$0$|$0$|$-355$

\endtable

There are nongeometric phases obtained by flopping the Kahler parameters.
One would need to analytically continue the expansions \fotwo\ to
calculate in these phases.  The boundary conditions on the disk
prevent the open instanton moduli space from being an
orbifold even when the closed string instanton moduli space is an orbifold
unless the boundary of the disk is fixed by the orbifold.  In this case 
fractional invariants are possible.

\bigskip\centerline{\bf Acknowledgments}\nobreak
I wish to acknowledge J. Distler and A. Iqbal for beneficial discussions
and R. McNees for answering Mathematica related questions.  This work
was supported in part by NSF grant PHY-0071512.

\footatend\vfill\supereject\immediate\closeout\rfile\writestoppt
\baselineskip=14pt\centerline{{\bf References}}\bigskip{\frenchspacing%
\parindent=20pt\escapechar=` \input refs.tmp\vfill\eject}\nonfrenchspacing

\end

%% file: onetime.tex
%
%
%
\def\unredoffs{} \def\redoffs{\voffset=-.31truein\hoffset=-.59truein}
\def\speclscape{\special{ps: landscape}}
%
%
%
%
\newbox\leftpage \newdimen\fullhsize \newdimen\hstitle \newdimen\hsbody
\tolerance=1000\hfuzz=2pt
\catcode`\@=11 
\def\bigans{b }
\message{ big or little (b/l)? }\read-1 to\answ
\ifx\answ\bigans\message{(This will come out unreduced.}
\magnification=1200\unredoffs\baselineskip=16pt plus 2pt minus 1pt
\hsbody=\hsize \hstitle=\hsize 
\else\message{(This will be reduced.} \let\l@r=L
\magnification=1000\baselineskip=16pt plus 2pt minus 1pt \vsize=7truein
\redoffs \hstitle=8truein\hsbody=4.75truein\fullhsize=10truein\hsize=\hsbody
\output={\ifnum\pageno=0 
  \shipout\vbox{\speclscape{\hsize\fullhsize\makeheadline}
    \hbox to \fullhsize{\hfill\pagebody\hfill}}\advancepageno
  \else
  \almostshipout{\leftline{\vbox{\pagebody\makefootline}}}\advancepageno 
  \fi}
\def\almostshipout#1{\if L\l@r \count1=1 \message{[\the\count0.\the\count1]}
      \global\setbox\leftpage=#1 \global\let\l@r=R
 \else \count1=2
  \shipout\vbox{\speclscape{\hsize\fullhsize\makeheadline}
      \hbox to\fullhsize{\box\leftpage\hfil#1}}  \global\let\l@r=L\fi}
\fi
%
\newcount\yearltd\yearltd=\year\advance\yearltd by -1900

\def\Title#1#2{\nopagenumbers\abstractfont\hsize=\hstitle\rightline{#1}%
\vskip 1in\centerline{\titlefont #2}\abstractfont\vskip .5in\pageno=0}
\def\Date#1{\vfill\leftline{#1}\tenpoint\supereject\global\hsize=\hsbody%
\footline={\nopagenumbers}}
%

\def\draftmode{\message{ DRAFTMODE }\def\draftdate{{\rm preliminary draft:
\number\month/\number\day/\number\yearltd\ \ \hourmin}}%
\headline={\hfil\draftdate}\writelabels\baselineskip=20pt plus 2pt minus 2pt
 {\count255=\time\divide\count255 by 60 \xdef\hourmin{\number\count255}
  \multiply\count255 by-60\advance\count255 by\time
  \xdef\hourmin{\hourmin:\ifnum\count255<10 0\fi\the\count255}}}
\def\nolabels{\def\wrlabeL##1{}\def\eqlabeL##1{}\def\reflabeL##1{}}
\def\writelabels{\def\wrlabeL##1{\leavevmode\vadjust{\rlap{\smash%
{\line{{\escapechar=` \hfill\rlap{\sevenrm\hskip.03in\string##1}}}}}}}%
\def\eqlabeL##1{{\escapechar-1\rlap{\sevenrm\hskip.05in\string##1}}}%
\def\reflabeL##1{\noexpand\llap{\noexpand\sevenrm\string\string\string##1}}}
\nolabels
%
\global\newcount\secno \global\secno=0
\global\newcount\meqno \global\meqno=1
\def\newsec#1{\global\advance\secno by1\message{(\the\secno. #1)}
\global\subsecno=0\eqnres@t\noindent{\bf\the\secno. #1}
\writetoca{{\secsym} {#1}}\par\nobreak\medskip\nobreak}
\def\eqnres@t{\xdef\secsym{\the\secno.}\global\meqno=1\bigbreak\bigskip}
\def\sequentialequations{\def\eqnres@t{\bigbreak}}\xdef\secsym{}
\global\newcount\subsecno \global\subsecno=0
\def\subsec#1{\global\advance\subsecno by1\message{(\secsym\the\subsecno. #1)}
\ifnum\lastpenalty>9000\else\bigbreak\fi
\noindent{\it\secsym\the\subsecno. #1}\writetoca{\string\quad 
{\secsym\the\subsecno.} {#1}}\par\nobreak\medskip\nobreak}
\def\appendix#1#2{\global\meqno=1\global\subsecno=0\xdef\secsym{\hbox{#1.}}
\bigbreak\bigskip\noindent{\bf Appendix #1. #2}\message{(#1. #2)}
\writetoca{Appendix {#1.} {#2}}\par\nobreak\medskip\nobreak}
%
%
\def\eqnn#1{\xdef #1{(\secsym\the\meqno)}\writedef{#1\leftbracket#1}%
\global\advance\meqno by1\wrlabeL#1}
\def\eqna#1{\xdef #1##1{\hbox{$(\secsym\the\meqno##1)$}}
\writedef{#1\numbersign1\leftbracket#1{\numbersign1}}%
\global\advance\meqno by1\wrlabeL{#1$\{\}$}}
\def\eqn#1#2{\xdef #1{(\secsym\the\meqno)}\writedef{#1\leftbracket#1}%
\global\advance\meqno by1$$#2\eqno#1\eqlabeL#1$$}
%
\newskip\footskip\footskip14pt plus 1pt minus 1pt 
\def\footnotefont{\ninepoint}\def\f@t#1{\footnotefont #1\@foot}
\def\f@@t{\baselineskip\footskip\bgroup\footnotefont\aftergroup\@foot\let\next}
\setbox\strutbox=\hbox{\vrule height9.5pt depth4.5pt width0pt}
\global\newcount\ftno \global\ftno=0
\def\foot{\global\advance\ftno by1\footnote{$^{\the\ftno}$}}
%
\newwrite\ftfile   
\def\footend{\def\foot{\global\advance\ftno by1\chardef\wfile=\ftfile
$^{\the\ftno}$\ifnum\ftno=1\immediate\openout\ftfile=foots.tmp\fi%
\immediate\write\ftfile{\noexpand\smallskip%
\noexpand\item{f\the\ftno:\ }\pctsign}\findarg}%
\def\footatend{\vfill\eject\immediate\closeout\ftfile{\parindent=20pt
\centerline{\bf Footnotes}\nobreak\bigskip\input foots.tmp }}}
\def\footatend{}
%
%
\global\newcount\refno \global\refno=1
\newwrite\rfile
\def\ref{[\the\refno]\nref}
\def\nref#1{\xdef#1{[\the\refno]}\writedef{#1\leftbracket#1}%
\ifnum\refno=1\immediate\openout\rfile=refs.tmp\fi
\global\advance\refno by1\chardef\wfile=\rfile\immediate
\write\rfile{\noexpand\item{#1\ }\reflabeL{#1\hskip.31in}\pctsign}\findarg}
\def\findarg#1#{\begingroup\obeylines\newlinechar=`\^^M\pass@rg}
{\obeylines\gdef\pass@rg#1{\writ@line\relax #1^^M\hbox{}^^M}%
\gdef\writ@line#1^^M{\expandafter\toks0\expandafter{\striprel@x #1}%
\edef\next{\the\toks0}\ifx\next\em@rk\let\next=\endgroup\else\ifx\next\empty%
\else\immediate\write\wfile{\the\toks0}\fi\let\next=\writ@line\fi\next\relax}}
\def\striprel@x#1{} \def\em@rk{\hbox{}} 
\def\lref{\begingroup\obeylines\lr@f}
\def\lr@f#1#2{\gdef#1{\ref#1{#2}}\endgroup\unskip}

\def\addref#1{\immediate\write\rfile{\noexpand\item{}#1}} 
\def\startrefs#1{\immediate\openout\rfile=refs.tmp\refno=#1}
\def\xref{\expandafter\xr@f}\def\xr@f[#1]{#1}
\def\refs#1{\count255=1[\r@fs #1{\hbox{}}]}
\def\r@fs#1{\ifx\und@fined#1\message{reflabel \string#1 is undefined.}%
\nref#1{need to supply reference \string#1.}\fi%
\vphantom{\hphantom{#1}}\edef\next{#1}\ifx\next\em@rk\def\next{}%
\else\ifx\next#1\ifodd\count255\relax\xref#1\count255=0\fi%
\else#1\count255=1\fi\let\next=\r@fs\fi\next}
%

%
\newwrite\ffile\global\newcount\figno \global\figno=1
\def\fig{fig.~\the\figno\nfig}
\def\nfig#1{\xdef#1{fig.~\the\figno}%
\writedef{#1\leftbracket fig.\noexpand~\the\figno}%
\ifnum\figno=1\immediate\openout\ffile=figs.tmp\fi\chardef\wfile=\ffile%
\immediate\write\ffile{\noexpand\medskip\noexpand\item{Fig.\ \the\figno. }
\reflabeL{#1\hskip.55in}\pctsign}\global\advance\figno by1\findarg}
\def\xfig{\expandafter\xf@g}\def\xf@g fig.\penalty\@M\ {}
\def\figs#1{figs.~\f@gs #1{\hbox{}}}
\def\f@gs#1{\edef\next{#1}\ifx\next\em@rk\def\next{}\else
\ifx\next#1\xfig #1\else#1\fi\let\next=\f@gs\fi\next}
\newwrite\lfile
{\escapechar-1\xdef\pctsign{\string\%}\xdef\leftbracket{\string\{}
\xdef\rightbracket{\string\}}\xdef\numbersign{\string\#}}

\def\writestop{\def\writestoppt{\immediate\write\lfile{\string\pageno%
\the\pageno\string\startrefs\leftbracket\the\refno\rightbracket%
\string\def\string\secsym\leftbracket\secsym\rightbracket%
\string\secno\the\secno\string\meqno\the\meqno}\immediate\closeout\lfile}}
\def\writestoppt{}\def\writedef#1{}
\def\seclab#1{\xdef #1{\the\secno}\writedef{#1\leftbracket#1}\wrlabeL{#1=#1}}
\def\subseclab#1{\xdef #1{\secsym\the\subsecno}%
\writedef{#1\leftbracket#1}\wrlabeL{#1=#1}}
\newwrite\tfile \def\writetoca#1{}
\def\leaderfill{\leaders\hbox to 1em{\hss.\hss}\hfill}
\def\writetoc{\immediate\openout\tfile=toc.tmp 
   \def\writetoca##1{{\edef\next{\write\tfile{\noindent ##1 
   \string\leaderfill {\noexpand\number\pageno} \par}}\next}}}

%
\catcode`\@=12 
%
\edef\tfontsize{\ifx\answ\bigans scaled\magstep3\else scaled\magstep4\fi}
\font\titlerm=cmr10 \tfontsize \font\titlerms=cmr7 \tfontsize
\font\titlermss=cmr5 \tfontsize \font\titlei=cmmi10 \tfontsize
\font\titleis=cmmi7 \tfontsize \font\titleiss=cmmi5 \tfontsize
\font\titlesy=cmsy10 \tfontsize \font\titlesys=cmsy7 \tfontsize
\font\titlesyss=cmsy5 \tfontsize \font\titleit=cmti10 \tfontsize
\skewchar\titlei='177 \skewchar\titleis='177 \skewchar\titleiss='177
\skewchar\titlesy='60 \skewchar\titlesys='60 \skewchar\titlesyss='60
\def\titlefont{\def\rm{\fam0\titlerm}
\textfont0=\titlerm \scriptfont0=\titlerms \scriptscriptfont0=\titlermss
\textfont1=\titlei \scriptfont1=\titleis \scriptscriptfont1=\titleiss
\textfont2=\titlesy \scriptfont2=\titlesys \scriptscriptfont2=\titlesyss
\textfont\itfam=\titleit \def\it{\fam\itfam\titleit}\rm}
 \ifx\answ\bigans\else scaled\magstep1\fi
\ifx\answ\bigans\def\abstractfont{\tenpoint}\else
\font\abssl=cmsl10 scaled \magstep1
\font\absrm=cmr10 scaled\magstep1 \font\absrms=cmr7 scaled\magstep1
\font\absrmss=cmr5 scaled\magstep1 \font\absi=cmmi10 scaled\magstep1
\font\absis=cmmi7 scaled\magstep1 \font\absiss=cmmi5 scaled\magstep1
\font\abssy=cmsy10 scaled\magstep1 \font\abssys=cmsy7 scaled\magstep1
\font\abssyss=cmsy5 scaled\magstep1 \font\absbf=cmbx10 scaled\magstep1
\skewchar\absi='177 \skewchar\absis='177 \skewchar\absiss='177
\skewchar\abssy='60 \skewchar\abssys='60 \skewchar\abssyss='60
\def\abstractfont{\def\rm{\fam0\absrm}
\textfont0=\absrm \scriptfont0=\absrms \scriptscriptfont0=\absrmss
\textfont1=\absi \scriptfont1=\absis \scriptscriptfont1=\absiss
\textfont2=\abssy \scriptfont2=\abssys \scriptscriptfont2=\abssyss
\textfont\itfam=\bigit \def\it{\fam\itfam\bigit}\def\footnotefont{\tenpoint}%
\textfont\slfam=\abssl \def\sl{\fam\slfam\abssl}%
\textfont\bffam=\absbf \def\bf{\fam\bffam\absbf}\rm}\fi
\def\tenpoint{\def\rm{\fam0\tenrm}
\textfont0=\tenrm \scriptfont0=\sevenrm \scriptscriptfont0=\fiverm
\textfont1=\teni  \scriptfont1=\seveni  \scriptscriptfont1=\fivei
\textfont2=\tensy \scriptfont2=\sevensy \scriptscriptfont2=\fivesy
\textfont\itfam=\tenit \def\it{\fam\itfam\tenit}\def\footnotefont{\ninepoint}%
\textfont\bffam=\tenbf \def\bf{\fam\bffam\tenbf}\def\sl{\fam\slfam\tensl}\rm}
\font\ninerm=cmr9 \font\sixrm=cmr6 \font\ninei=cmmi9 \font\sixi=cmmi6 
\font\ninesy=cmsy9 \font\sixsy=cmsy6 \font\ninebf=cmbx9 
\font\nineit=cmti9 \font\ninesl=cmsl9 \skewchar\ninei='177
\skewchar\sixi='177 \skewchar\ninesy='60 \skewchar\sixsy='60 
\def\ninepoint{\def\rm{\fam0\ninerm}
\textfont0=\ninerm \scriptfont0=\sixrm \scriptscriptfont0=\fiverm
\textfont1=\ninei \scriptfont1=\sixi \scriptscriptfont1=\fivei
\textfont2=\ninesy \scriptfont2=\sixsy \scriptscriptfont2=\fivesy
\textfont\itfam=\ninei \def\it{\fam\itfam\nineit}\def\sl{\fam\slfam\ninesl}%
\textfont\bffam=\ninebf \def\bf{\fam\bffam\ninebf}\rm} 
%
%
\def\noblackbox{\overfullrule=0pt}
\hyphenation{anom-aly anom-alies coun-ter-term coun-ter-terms}
\def\inv{^{\raise.15ex\hbox{${\scriptscriptstyle -}$}\kern-.05em 1}}

\def\Dsl{\,\raise.15ex\hbox{/}\mkern-13.5mu D} 
\def\dsl{\raise.15ex\hbox{/}\kern-.57em\partial}

\font\bigit=cmti10 scaled \magstep1
\def\lspace{\ifx\answ\bigans{}\else\qquad\fi}
\def\lbspace{\ifx\answ\bigans{}\else\hskip-.2in\fi} 
\def\boxeqn#1{\vcenter{\vbox{\hrule\hbox{\vrule\kern3pt\vbox{\kern3pt
	\hbox{${\displaystyle #1}$}\kern3pt}\kern3pt\vrule}\hrule}}}
\def\mbox#1#2{\vcenter{\hrule \hbox{\vrule height#2in
		\kern#1in \vrule} \hrule}}  
%

\def\darr#1{\raise1.5ex\hbox{$\leftrightarrow$}\mkern-16.5mu #1}

\def\roughly#1{\raise.3ex\hbox{$#1$\kern-.75em\lower1ex\hbox{$\sim$}}}

%% file: tables.tex
%
\newbox\hdbox%
\newcount\hdrows%
\newcount\multispancount%
\newcount\ncase%
\newcount\ncols
\newcount\nrows%
\newcount\nspan%
\newcount\ntemp%
\newdimen\hdsize%
\newdimen\newhdsize%
\newdimen\parasize%
\newdimen\spreadwidth%
\newdimen\thicksize%
\newdimen\thinsize%
\newdimen\tablewidth%
\newif\ifcentertables%
\newif\ifendsize%
\newif\iffirstrow%
\newif\iftableinfo%
\newtoks\dbt%
\newtoks\hdtks%
\newtoks\savetks%
\newtoks\tableLETtokens%
\newtoks\tabletokens%
\newtoks\widthspec%
%
%
\immediate\write15{%
CP SMSG GJMSINK TEXTABLE --> TABLE MACROS V. 851121 JOB = \jobname%
}%
%
%
\tableinfotrue%
\catcode`\@=11
%
%
\def\tstrut{\vrule height3.1ex depth1.2ex width0pt}%
\def\and{\char`\&}
\def\tablerule{\noalign{\hrule height\thinsize depth0pt}}%
\thicksize=1.5pt
\thinsize=0.6pt
\def\thickrule{\noalign{\hrule height\thicksize depth0pt}}%
\def\ctr#1{\hfil\ #1\hfil}%
%
%
%
%
\tablewidth=-\maxdimen%
\spreadwidth=-\maxdimen%
\def\tabskipglue{0pt plus 1fil minus 1fil}%
%
%
\centertablestrue%
%
%
%
%
\parasize=4in%
\gdef\ARGS{########}
\gdef\headerARGS{####}
\def\@mpersand{&}
{\catcode`\|=13
\gdef\letbarzero{\let|0}
\gdef\letbartab{\def|{&&}}%
\gdef\letvbbar{\let\vb|}%
}
{\catcode`\&=4
\def\ampskip{&\omit\hfil&}
\catcode`\&=13
\let&0
\xdef\letampskip{\def&{\ampskip}}%
\gdef\letnovbamp{\let\novb&\let\tab&}
}
\def\begintable{
   \begingroup%
   \catcode`\|=13\letbartab\letvbbar%
   \catcode`\&=13\letampskip\letnovbamp%
   \def\multispan##1{
      \omit \mscount##1%
      \multiply\mscount\tw@\advance\mscount\m@ne%
      \loop\ifnum\mscount>\@ne \sp@n\repeat%
   }
   \def\|{%
      &\omit\widevline&%
   }%
   \ruledtable
}
\long\def\ruledtable#1\endtable{%
%
%
%
   \offinterlineskip
   \tabskip 0pt
   \def\widevline{\vrule width\thicksize}
   \def\endrow{\@mpersand\omit\hfil\crnorm\@mpersand}%
   \def\crthick{\@mpersand\crnorm\thickrule\@mpersand}%
   \def\crthickneg##1{\@mpersand\crnorm\thickrule
          \noalign{{\skip0=##1\vskip-\skip0}}\@mpersand}%
   \def\crnorule{\@mpersand\crnorm\@mpersand}%
   \def\crnoruleneg##1{\@mpersand\crnorm
          \noalign{{\skip0=##1\vskip-\skip0}}\@mpersand}%
   \let\nr=\crnorule
   \def\endtable{\@mpersand\crnorm\thickrule}%
   \let\crnorm=\cr
%
%
   \edef\cr{\@mpersand\crnorm\tablerule\@mpersand}%
   \def\crneg##1{\@mpersand\crnorm\tablerule
          \noalign{{\skip0=##1\vskip-\skip0}}\@mpersand}%
   \let\ctneg=\crthickneg
   \let\nrneg=\crnoruleneg
   \the\tableLETtokens
%
%
   \tabletokens={&#1}
%
%
   \countROWS\tabletokens\into\nrows%
   \countCOLS\tabletokens\into\ncols%
%
%
   \advance\ncols by -1%
   \divide\ncols by 2%
   \advance\nrows by 1%
%
%
   \iftableinfo %
      \immediate\write16{[Nrows=\the\nrows, Ncols=\the\ncols]}%
   \fi%
%
%
   \ifcentertables
      \ifhmode \par\fi
      \line{
      \hss
   \else %
      \hbox{%
   \fi
      \vbox{%
         \makePREAMBLE{\the\ncols}
         \edef\next{\preamble}
         \let\preamble=\next
         \makeTABLE{\preamble}{\tabletokens}
      }
      \ifcentertables \hss}\else }\fi
   \endgroup
   \tablewidth=-\maxdimen
   \spreadwidth=-\maxdimen
}
\def\makeTABLE#1#2{
   {
   \let\ifmath0
   \let\header0
   \let\multispan0
%
%
   \ncase=0%
   \ifdim\tablewidth>-\maxdimen \ncase=1\fi%
   \ifdim\spreadwidth>-\maxdimen \ncase=2\fi%
   \relax
%
   \ifcase\ncase %
      \widthspec={}%
   \or %
      \widthspec=\expandafter{\expandafter t\expandafter o%
                 \the\tablewidth}%
   \else %
      \widthspec=\expandafter{\expandafter s\expandafter p\expandafter r%
                 \expandafter e\expandafter a\expandafter d%
                 \the\spreadwidth}%
   \fi %
   \xdef\next{
      \halign\the\widthspec{%
      #1
      \noalign{\hrule height\thicksize depth0pt}
      \the#2\endtable
%
      }
   }
   }
   \next
}
\def\makePREAMBLE#1{
   \ncols=#1
   \begingroup
   \let\ARGS=0
   \edef\xtp{\widevline\ARGS\tabskip\tabskipglue%
   &\ctr{\ARGS}\tstrut}
   \advance\ncols by -1
   \loop
      \ifnum\ncols>0 %
      \advance\ncols by -1%
      \edef\xtp{\xtp&\vrule width\thinsize\ARGS&\ctr{\ARGS}}%
   \repeat
   \xdef\preamble{\xtp&\widevline\ARGS\tabskip0pt%
   \crnorm}
   \endgroup
}
\def\countROWS#1\into#2{
   \let\countREGISTER=#2%
   \countREGISTER=0%
   \expandafter\ROWcount\the#1\endcount%
}%
\def\ROWcount{%
   \afterassignment\subROWcount\let\next= %
}%
\def\subROWcount{%
   \ifx\next\endcount %
      \let\next=\relax%
   \else%
      \ncase=0%
      \ifx\next\cr %
         \global\advance\countREGISTER by 1%
         \ncase=0%
      \fi%
      \ifx\next\endrow %
         \global\advance\countREGISTER by 1%
         \ncase=0%
      \fi%
      \ifx\next\crthick %
         \global\advance\countREGISTER by 1%
         \ncase=0%
      \fi%
      \ifx\next\crnorule %
         \global\advance\countREGISTER by 1%
         \ncase=0%
      \fi%
      \ifx\next\crthickneg %
         \global\advance\countREGISTER by 1%
         \ncase=0%
      \fi%
      \ifx\next\crnoruleneg %
         \global\advance\countREGISTER by 1%
         \ncase=0%
      \fi%
      \ifx\next\crneg %
         \global\advance\countREGISTER by 1%
         \ncase=0%
      \fi%
      \ifx\next\header %
         \ncase=1%
      \fi%
      \relax%
      \ifcase\ncase %
         \let\next\ROWcount%
      \or %
         \let\next\argROWskip%
      \else %
      \fi%
   \fi%
   \next%
}
\def\counthdROWS#1\into#2{%
\dvr{10}%
   \let\countREGISTER=#2%
   \countREGISTER=0%
\dvr{11}%
\dvr{13}%
   \expandafter\hdROWcount\the#1\endcount%
\dvr{12}%
}%
\def\hdROWcount{%
   \afterassignment\subhdROWcount\let\next= %
}%
\def\subhdROWcount{%
   \ifx\next\endcount %
      \let\next=\relax%
   \else%
      \ncase=0%
      \ifx\next\cr %
         \global\advance\countREGISTER by 1%
         \ncase=0%
      \fi%
      \ifx\next\endrow %
         \global\advance\countREGISTER by 1%
         \ncase=0%
      \fi%
      \ifx\next\crthick %
         \global\advance\countREGISTER by 1%
         \ncase=0%
      \fi%
      \ifx\next\crnorule %
         \global\advance\countREGISTER by 1%
         \ncase=0%
      \fi%
      \ifx\next\header %
         \ncase=1%
      \fi%
\relax%
      \ifcase\ncase %
         \let\next\hdROWcount%
      \or%
         \let\next\arghdROWskip%
      \else %
      \fi%
   \fi%
   \next%
}%
{\catcode`\|=13\letbartab
\gdef\countCOLS#1\into#2{%
   \let\countREGISTER=#2%
   \global\countREGISTER=0%
   \global\multispancount=0%
   \global\firstrowtrue
   \expandafter\COLcount\the#1\endcount%
   \global\advance\countREGISTER by 3%
   \global\advance\countREGISTER by -\multispancount
}%
\gdef\COLcount{%
   \afterassignment\subCOLcount\let\next= %
}%
{\catcode`\&=13%
\gdef\subCOLcount{%
   \ifx\next\endcount %
      \let\next=\relax%
   \else%
      \ncase=0%
      \iffirstrow
         \ifx\next& %
            \global\advance\countREGISTER by 2%
            \ncase=0%
         \fi%
         \ifx\next\span %
            \global\advance\countREGISTER by 1%
            \ncase=0%
         \fi%
         \ifx\next| %
            \global\advance\countREGISTER by 2%
            \ncase=0%
         \fi
         \ifx\next\|
            \global\advance\countREGISTER by 2%
            \ncase=0%
         \fi
         \ifx\next\multispan
            \ncase=1%
            \global\advance\multispancount by 1%
         \fi
         \ifx\next\header
            \ncase=2%
         \fi
         \ifx\next\cr       \global\firstrowfalse \fi
         \ifx\next\endrow   \global\firstrowfalse \fi
         \ifx\next\crthick  \global\firstrowfalse \fi
         \ifx\next\crnorule \global\firstrowfalse \fi
         \ifx\next\crnoruleneg \global\firstrowfalse \fi
         \ifx\next\crthickneg  \global\firstrowfalse \fi
         \ifx\next\crneg       \global\firstrowfalse \fi
      \fi
\relax
      \ifcase\ncase %
         \let\next\COLcount%
      \or %
         \let\next\spancount%
      \or %
         \let\next\argCOLskip%
      \else %
      \fi %
   \fi%
   \next%
}%
\gdef\argROWskip#1{%
   \let\next\ROWcount \next%
}
\gdef\arghdROWskip#1{%
   \let\next\ROWcount \next%
}
\gdef\argCOLskip#1{%
   \let\next\COLcount \next%
}
}
}
\def\spancount#1{
   \nspan=#1\multiply\nspan by 2\advance\nspan by -1%
   \global\advance \countREGISTER by \nspan
   \let\next\COLcount \next}%
\def\dvr#1{\relax}%
\def\header#1{%
\dvr{1}{\let\cr=\@mpersand%
\hdtks={#1}%
\counthdROWS\hdtks\into\hdrows%
\advance\hdrows by 1%
\ifnum\hdrows=0 \hdrows=1 \fi%
\dvr{5}\makehdPREAMBLE{\the\hdrows}%
\dvr{6}\getHDdimen{#1}%
{\parindent=0pt\hsize=\hdsize{\let\ifmath0%
\xdef\next{\valign{\headerpreamble #1\crnorm}}}\dvr{7}\next\dvr{8}%
}%
}\dvr{2}}
\def\makehdPREAMBLE#1{
\dvr{3}%
\hdrows=#1
{
\let\headerARGS=0%
\let\cr=\crnorm%
\edef\xtp{\vfil\hfil\hbox{\headerARGS}\hfil\vfil}%
\advance\hdrows by -1
\loop
\ifnum\hdrows>0%
\advance\hdrows by -1%
\edef\xtp{\xtp&\vfil\hfil\hbox{\headerARGS}\hfil\vfil}%
\repeat%
\xdef\headerpreamble{\xtp\crcr}%
}
\dvr{4}}
\def\getHDdimen#1{%
\hdsize=0pt%
\getsize#1\cr\end\cr%
}
\def\getsize#1\cr{%
\endsizefalse\savetks={#1}%
\expandafter\lookend\the\savetks\cr%
\relax \ifendsize \let\next\relax \else%
\setbox\hdbox=\hbox{#1}\newhdsize=1.0\wd\hdbox%
\ifdim\newhdsize>\hdsize \hdsize=\newhdsize \fi%
\let\next\getsize \fi%
\next%
}%
\def\lookend{\afterassignment\sublookend\let\looknext= }%
\def\sublookend{\relax%
\ifx\looknext\cr %
\let\looknext\relax \else %
   \relax
   \ifx\looknext\end \global\endsizetrue \fi%
   \let\looknext=\lookend%
    \fi \looknext%
}%
%
%
\def\tablelet#1{%
   \tableLETtokens=\expandafter{\the\tableLETtokens #1}%
}%
\catcode`\@=12